\documentclass[preprint]{revtex4-2}

\usepackage{graphicx}
\usepackage{siunitx}
\usepackage[hidelinks]{hyperref}
\usepackage{gensymb}
\usepackage{float}
\usepackage{lineno}
\usepackage{xcolor}
\usepackage{amsmath, amsthm, amssymb, bm, bbm}

\newcommand{\angstrom}{\text{\normalfont\AA}}
\graphicspath{{figures/}}

\newcommand{\mexpect}{\mathop{{}\mathbb{E}}{}}

\newcommand{\eeff}{\epsilon_{\text{eff}}}
\newcommand{\newmid }{|}
\renewcommand{\paragraph}[1]{\textbf{#1}.\quad }

\begin{document}

\title{Fast and Reliable Probabilistic Reflectometry Inversion with Prior-Amortized Neural Posterior Estimation }

\author{Vladimir Starostin\textsuperscript{a}}
\author{Maximilian Dax\textsuperscript{b}}
\author{Alexander Gerlach\textsuperscript{a}}
\author{Alexander Hinderhofer\textsuperscript{a}}
\author{Álvaro Tejero-Cantero\textsuperscript{a}}
\author{Frank Schreiber\textsuperscript{a}}

\affiliation{
[a] University of Tübingen, 72076 Tübingen, Germany\\
[b] Max Planck Institute for Intelligent Systems, 72076 Tübingen, Germany
}

\begin{abstract} 

    Reconstructing the structure of thin films and multilayers from measurements of scattered X-rays or neutrons is key to progress in physics, chemistry, and biology. However, finding all structures compatible with reflectometry data is computationally prohibitive for standard algorithms, which typically results in unreliable analysis with only a single potential solution identified. We address this lack of reliability with a probabilistic deep learning method that identifies all realistic structures in seconds, setting new standards in reflectometry. Our method, Prior-Amortized Neural Posterior Estimation (PANPE), combines simulation-based inference with novel adaptive priors that inform the inference network about known structural properties and controllable experimental conditions. PANPE networks support key scenarios such as high-throughput sample characterization, real-time monitoring of evolving structures, or the co-refinement of several experimental data sets, and can be adapted to provide fast, reliable, and flexible inference across many other inverse problems.  
    
\end{abstract}

\maketitle
\renewcommand\thefigure{\arabic{figure}}
\renewcommand\thetable{\arabic{table}}

\section{Introduction}

    Scattering techniques enable the reconstruction of object structures through the analysis of scattered radiation \cite{AlsNielsen2011Elements, Feidenhansl1989}. At the nanoscale, this requires radiation with short wavelengths, such as X-rays and thermal neutrons. While for the reconstruction of images from visible scattered light there are more established tools including optical lenses, employing these tools for X-rays and neutrons frequently poses significant challenges, leading to the use of algorithms for the reconstruction process \cite{Fienup1982}. These algorithms, however, receive incomplete information, as detectors capture intensities, but not the phase information of the scattered waves. This gives rise to the phaseless inverse problem in scattering physics. While physical models can simulate scattered intensities from a given structure, reconstructing the structure from actual measurements is analytically intractable, and experimental data can be consistent with multiple physical structures \cite{Bertero2024-dz}. This ambiguity can be then resolved through complementary measurements or physical knowledge, but it is crucial to first acknowledge the existence of multiple solutions to avoid costly misinterpretations of the data. Together with advances in experimental methods enabling time-resolved online experiments and high-throughput pipelines \cite{Wang2018, SchumiMareek2024}, this creates a pressing need for algorithms that are (1) fast, (2) capable of reliably identifying all possible solutions, and (3) flexible enough to integrate additional data and physics-informed constraints.

\begin{figure}[h!]
        \includegraphics[width=0.64 \linewidth]{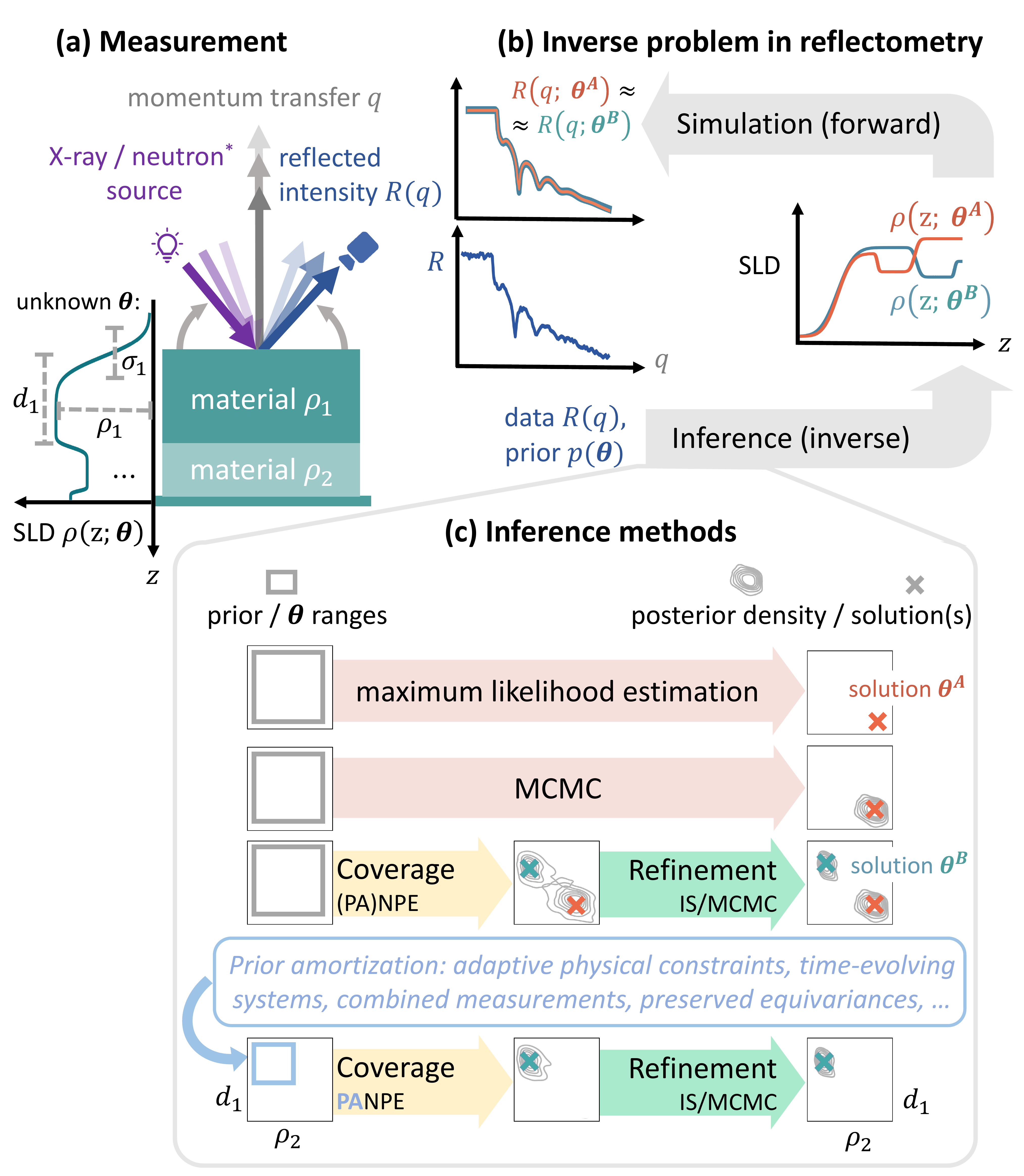}
        \caption{Reflectometry analysis. (a) A schematic experimental setup for reflectometry measurements. The reflected intensity $R(q)$ from a studied layered structure as a function of momentum transfer $q$ contains information about parameters $\theta$ of the studied sample. The momentum transfer is typically controlled by the geometry in X-ray reflectometry or by the energy in time-of-flight neutron measurements. (b) Inverse problem in reflectometry: the forward simulations provided by the scattering theory should be inverted during inference, which is generally ambiguous. (c) Inference methods commonly employed for reflectometry analysis, as well as our proposed approach. The standard maximum likelihood estimation approach provides a single solution by design. MCMC locally explores the parameter space and can overlook distributional modes. In contrast, (PA)NPE posterior estimate is guaranteed to cover all the solutions, with further refinement based on likelihood evaluation improving accuracy. Our prior amortization method, PANPE, enables the analysis of multiple experimental scenarios using a single neural network.}

        \label{fig:pipeline}
    \end{figure}

    The need of fast and reliable algorithms is especially evident for neutron and X-ray reflectometry \cite{Parratt1954Surface, PhysRevB.38.2297}. The reflected intensity $R$ in specular geometry as a function of momentum transfer $q$ (see \autoref{fig:pipeline}a) can inform about the scattering length density (SLD) profile for a broad range of thin films and layered structures, ranging from solar cells \cite{Brinkmann2022} to biological membranes \cite{Armanious2022, Caselli2024}. The SLD profile is typically modeled by parameters $\bm\theta$ that include layer thicknesses $d_l$, densities $\rho_l$, and interface roughnesses $\sigma_l$. Obtaining these parameters $\bm\theta$ from a reflectivity curve $R(q)$ in a fast and reliable way is a longstanding inverse problem (\autoref{fig:pipeline}b) due to the phase loss, measurement noise, and the limited range and resolution of $q$. For a long time, a common approach was to search for the ``best'' \textit{single} set of parameters $\bm\theta^*$ maximizing the likelihood of the measured data. However, maximum likelihood estimation remains fundamentally unreliable as it overlooks other potential physical solutions arising from ambiguity in the inverse problem. To address ambiguity, we need to embark on a principled probabilistic approach, and estimate the posterior probability density of the parameters $\bm\theta$ given the measured data $\bm{R}$. In such a Bayesian posterior $p(\bm\theta | \bm{R})$, different probable structures appear as distributional modes (\autoref{fig:pipeline}c). In practice, the inference of a high-dimensional posterior is inherently challenging, and particularly so in reflectometry where multiple narrow distributional modes are common. Conventional Bayesian likelihood-based techniques such as Markov Chain Monte Carlo (MCMC)~\cite{Metropolis1953} are neither fast nor reliable, as they generally miss distributional modes.

    Here, we present a machine learning solution for Bayesian reflectometry analysis  that provides fast, reliable and accurate inference along with the flexibility that online experiments demand. \textbf{Speed} is achieved by pre-training a neural network across large amounts of representative data, an \textit{amortization} procedure that then allows for real-time inference on new samples. \textbf{Reliability} stems from the use of recent simulation-based inference which, in contrast to likelihood-based modes provides comprehensive coverage of the search space. \textbf{Accuracy} in the identified solutions is achieved via a subsequent likelihood-based step, that refines machine learning-based estimates. Fast likelihood evaluations are possible thanks to a novel GPU-parallel transfer-matrix simulator. Finally, we enable great \textbf{flexibility} for a wide range of standard scenarios in experimental setups by extending neural posterior estimation (NPE) with prior amortization, which we term Prior-Amortized Neural Posterior Estimation (PANPE). Our method PANPE can use dynamically set, adaptive prior distributions, allowing to track online expriments, leverages equivariance transformations to enable amortization over different $q$ ranges, and can combine evidence from multiple measurements for efficient inference. Below, we describe in detail how PANPE works and benchmark its performance on real and synthetic reflectometry data.

\section{Results}

    We describe our method in Section \ref{sec:results:overview}. We then evaluate it on synthetic data, which encompasses a broad range of structures, in Section \ref{sec:results:sim}. Subsequently, we apply it to X-ray reflectometry (XRR) data from online \textit{in situ} experiments in Section \ref{sec:results:exp-xrr}. Finally, we showcase our approach for combining multiple measurements in neutron reflectometry (NR) data in Section \ref{sec:results:neutron}.

\subsection{Overview of PANPE}\label{sec:results:overview}

    \paragraph{Bayesian framework for inverse problems} Reflectometry analysis aims to infer physical parameters $\bm\theta$ from measured data $\bm{R}$. Each parameter set $\bm\theta$ describes a hypothetical SLD profile of the studied structure (SLD parameterization is discussed in Methods \ref{sec:methods:theta}). For a given reflectometry measurement $\bm{R}$, the Bayesian posterior distribution~\cite{bayes1763lii}
    \begin{equation}\label{eq:bayes1}
        p(\bm{\theta} \newmid \bm{R}) \propto p(\bm{R} \newmid \bm{\theta})p(\bm{\theta})
    \end{equation}
    offers a probabilistic estimate of $\bm{\theta}$, characterized by the likelihood $p(\bm{R} \newmid \bm{\theta})$ provided by scattering theory and a prior $p(\bm{\theta})$ provided by experimentalists. 

    Importantly, the prior physical knowledge about the studied structure, formulated as a prior distribution $p(\bm\theta)$ over parameters $\bm\theta$ in Bayesian framework, serves as a crucial tool for resolving ambiguity in reflectometry and, more broadly, in scattering physics. It facilitates physics-informed analysis by integrating knowledge about the materials used and other properties of the system under study. Indeed, what may be unphysical scattering length density (SLD) profiles in one context can be considered legitimate solutions in another, depending on the known properties of the system being investigated. 

    With Bayes's theorem, we can use likelihood and prior to calculate the (unnormalized) posterior density for any given parameter set $\bm\theta_i$. However, such density evaluation does not directly provide most practically relevant estimates, such as mean values or confidence intervals. To compute them, we need to \textit{draw samples} $\{\bm\theta_i\}_{i=1}^N \sim p(\bm\theta \newmid \bm{R})$ from the posterior distribution, i.e. randomly selecting parameter sets $\bm\theta_i$ in proportion to their posterior density $p(\bm\theta_i \newmid \bm{R})$. The complexity of this operation for high-dimensional parameters $\bm\theta$ renders Bayesian inference a highly challenging task, traditionally limiting the applicability of the Bayesian approach to simple cases.

    \paragraph{Coverage and refinement of Neural Posterior Estimation} To reliably sample from the Bayesian posterior, we adopt a recent Neural Posterior Estimation (NPE) method \cite{Cranmer2020, normflows2021Papamakarios}. It employs a normalizing flow architecture~\cite{rezende2015variational, neuralsplineflows2019, normflows2021Papamakarios} as the neural network-based density estimator. Normalizing flows can learn complex high-dimensional conditional distributions and have been employed for Bayesian inference in multiple applications. Once trained on a broad range of simulated data, the flow-based model $p_{\mathrm{NN}}(\bm\theta \newmid \bm{R})$ can efficiently generate samples $\{\bm\theta_i\}_{i=1}^N \sim p_{\mathrm{NN}}(\bm\theta \newmid \bm{R})$ and evaluate densities $p_i = p_{\mathrm{NN}}(\bm\theta_i \newmid \bm{R})$ for different measurements $\bm{R}$. 

     A key property of normalizing flows is that the exact density evaluation enables training the model by minimizing the \textit{forward} Kullback–Leibler (KL) divergence (see Methods \ref{sec:methods:panpe}). This ensures the \textit{coverage} property of NPE, i.e. it contains the full support of the true unknown posterior $p(\bm\theta \newmid \bm{R})$ and does not miss distributional modes~\cite{Dax2023NIS}. However, the ``shape'' of the NN-based density estimate might deviate from the target posterior. To address this, likelihood-based methods such as IS and MCMC can refine the NPE results for more accurate estimates. Thus, NPE ensures the coverage property, while likelihood-based refinement enhances accuracy.

    Our custom-made GPU-accelerated transfer-matrix reflectometry simulator~\cite{Abeles1950La} implemented using PyTorch~\cite{pytorch} accelerates both the training and inference stages. This allows us to simulate new curves directly during the training process without reusing simulations, thereby preventing overfitting. Furthermore, as shown in Section~\ref{sec:results:sim}, our reflectometry analysis with importance sampling refinement -- which requires likelihood evaluations -- typically takes just seconds to less than a minute on a single graphics card, the NVIDIA RTX 2080 Ti.
    
     \paragraph{Amortization across various experimental scenarios} Along with reliability due to the coverage property, NPE provides fast inference, since its computational cost is \textit{amortized} by the training process. However, the amortization also introduces a key practical limitation, as a trained model can only operate within some predefined training ranges of parameters. In the case of reflectometry, this limitation not only includes the parameter ranges, but also experimental settings such as the discretization of the momentum transfer axis $q$ or the measurement uncertainties, all of which may significantly affect the resulting posterior estimate. We address these limitations by implementing intensive amortization, taking into account the diverse varying aspects of the experiment, such as $q$ discretization and range, measurement uncertainties, and prior information.
    
    Measurements of the reflected intensity $R(q)$ are taken at different discrete values of momentum transfer $q$ (see \autoref{fig:pipeline}a). Together, these measurements form an input dataset $\bm{R} = \{q_p, R(q_p), s_p\}_{p=1}^{n_q}$ consisting of $n_q$ measured points (here $s_p$ represents the uncertainty of each data point). Our approach accommodates experiments featuring arbitrarily spaced $q$ points and varying numbers of measurements $n_q$, by utilizing an efficient embedding neural network equipped with trainable interpolation kernels (see Methods \ref{sec:methods:inference}).

     However, in the context of reflectometry analysis, the most significant variable component is the prior information. Indeed, known constraints on physical properties vary substantially across different structures under study. Furthermore, in the online experiments discussed in Section \ref{sec:results:exp-xrr}, the experimentalists may modify the structure by changing control parameters -- and accordingly, the respective priors. Adjustments to the priors are also necessary when combining multiple measurements in neutron reflectometry, as considered in Section \ref{sec:results:neutron}. In these and other scenarios, the analysis must adapt to the changing prior distribution. Standard machine learning solutions like NPE, which typically assume a fixed prior distribution, fall short under these experimental conditions. To overcome this limitation, we introduce Prior-Amortized Neural Posterior Estimation (PANPE) that accommodates a variety of prior distributions within a single model. 
     
    \paragraph{Prior-Amortized Neural Posterior Estimation (PANPE)} We incorporate dynamic prior information into a neural network by choosing a class of distributions $p(\bm\theta \newmid \bm\phi)$ parameterized by $\bm\phi$. The newly-introduced parameters $\bm\phi$ reflect prior information about the system. They are supplied as an additional input to the flow-based neural network $p_{\mathrm{NN}}(\bm\theta | \bm{R}, \bm\phi)$ alongside the measured data. This allows us to train a single neural network and amortize inference across both measurements $\bm{R}$ \textit{and} priors $p(\bm\theta \newmid \bm\phi)$:

    \begin{equation}\label{eq:bayes2}
        p_{\mathrm{NN}}(\bm\theta | \bm{R}, \bm\phi) \approx p(\bm\theta \newmid \bm{R}, \bm\phi) \propto p(\bm{R} \newmid \bm\theta) p(\bm\theta \newmid \bm\phi)  \text{ . }
    \end{equation}

    In reflectometry analysis, it is typically sufficient to employ uniform prior distributions $p(\bm\theta) = \prod^{n}_{j=1} U(\theta_j^{min}, \theta_j^{max})$, where $n$ is the number of parameters $\bm\theta$. Thus, we define $\bm\phi$ as a set of corresponding parameter ranges: $\bm\phi = \{\theta^{\mathrm{min}}_j, \theta^{\mathrm{max}}_j\}_{j=1}^{n}$. This results in $2n = 20$ additional input values for a task with $n = 10$ parameters $\bm\theta$. In this manner, the model is trained to approximate posterior distribution for a continuous set $p(\bm\phi)$ of (uniform) prior distributions $p(\bm\theta \newmid \bm\phi)$ within a larger parameter space. Indeed, our approach can be extended to other classes of distributions by providing suitable parameterization. We discuss why the likelihood-based refinement (or rejection sampling in the case of uniform prior distribution) is not a practical alternative to the prior amortization in out case in Section \ref{sec:results:sim}. Our prior amortization approach is discussed in detail in Methods \ref{sec:methods:panpe}.

    We explain the inference pipeline in Methods \ref{sec:methods:inference}. Given the data $\bm{R}$ and the prior distribution characterized by parameters $\bm\phi$, we sample from the trained PANPE model and apply likelihood-based refinement either using importance sampling~\cite{Dax2023NIS} (PANPE-IS) or MCMC (PANPE-MCMC). 

    \paragraph{Parameter-conditioned posterior estimation} In certain cases, it is required to estimate parameter-conditioned posterior estimation, where instead of providing narrow priors for a parameter, it is fixed. We show one such case in Section \ref{sec:results:neutron} where parameter-conditioned posterior estimation is necessary for combining multiple measurements with partially shared parameters. 
    
    Another scenario relevant for future reflectometry applications considers choosing the appropriate physical model: by setting the thicknesses of a subset of layers to zero, one can effectively change the number of layers in the physical model. Consequently, several physical models represented by different numbers of free parameters can be compared via standard criteria such as the Bayes factor $\frac{p(\bm{R} \newmid \bm\phi_1)}{p(\bm{R} \newmid \bm\phi_2)}$ using the same neural network.
    
    Fixing some parameters indeed changes the dimensionality of the remaining free parameters, which is not supported by standard implementations of the normalizing flow. To circumvent this limitation, we introduce a reparameterization procedure of the parameter space that enables us to sample from the parameter-conditioned posterior estimation by providing ``zero-width'' prior. We discuss this approach in Methods \ref{sec:methods:panpe} and employ it in Section \ref{sec:results:neutron}.

    \paragraph{Preserved equivariances in the density estimator} The reflectometry simulator features a number of simple deterministic functional relationships between the input $\{q, \bm{\theta}\}$ and the simulated reflectivity curve $R(q, \bm{\theta})$. We systematically review them in Methods \ref{sec:methods:invariance}. For instance, these include the unit scaling invariance: rescaling the momentum transfer axis $\bm{q} \to u \cdot \bm{q}$ ($u \in \mathbb{R}_{> 0}$) together with a certain parameter rescaling transformation $\bm{\theta} \to T_u(\bm{\theta})$ does not alter the result: $R(u \cdot \bm{q}, T_u(\bm{\theta})) = R(\bm{q}, \bm{\theta})$. Conventional reflectometry analysis does not need to consider these relationships, but they become critical in amortized machine learning solutions: the trained model must reflect these relationships, ensuring that specific changes in the input data to the density estimator result in corresponding changes in the posterior distribution. To enhance the performance of the model, we directly incorporate these relationships into the inference pipeline (see Methods \ref{sec:methods:invariance} and 
    \ref{sec:methods:inference}), rather than having the model learn them from data. We note that the prior amortization is generally required for this operation, since the considered transformations alter the prior distribution.

    \paragraph{Related work} 
    Sequential NPE~\cite{papamakarios2016fast,lueckmann2017flexible,greenberg2019automatic,deistler2022truncated} effectively applies data-informed prior updates, but such methods require simulation and neural network training at inference time. The Simformer~\cite{gloeckler2024all} framework enables prior updates by leveraging diffusion guidance. Equivariances between parameter and data spaces can be integrated with group-equivariant NPE~\cite{dax2021real,Dax2022GNPE}, but this method requires iterative inference and is therby slower than NPE. The \textsc{Dingo-BNS} framework~\cite{dax2024real} for gravitational-wave inference combines adaptive priors with amortized NPE to achieve improved data compression. Our study builds on and extends upon these works, demonstrating how adaptive SBI priors enable posterior zooming into arbitrary parts of the parameter space in a high-profile science application.

\subsection{Performance on simulated data}\label{sec:results:sim}

        We first test our model on the test simulated dataset with a broad range of two-layer structures with 10 parameters. Performance on the X-ray reflectometry data is discussed in Section \ref{sec:results:exp-xrr}. For the results on neutron reflectometry (NR), refer to Section \ref{sec:results:neutron}.

        \begin{figure}
            \includegraphics[width=0.65\linewidth]{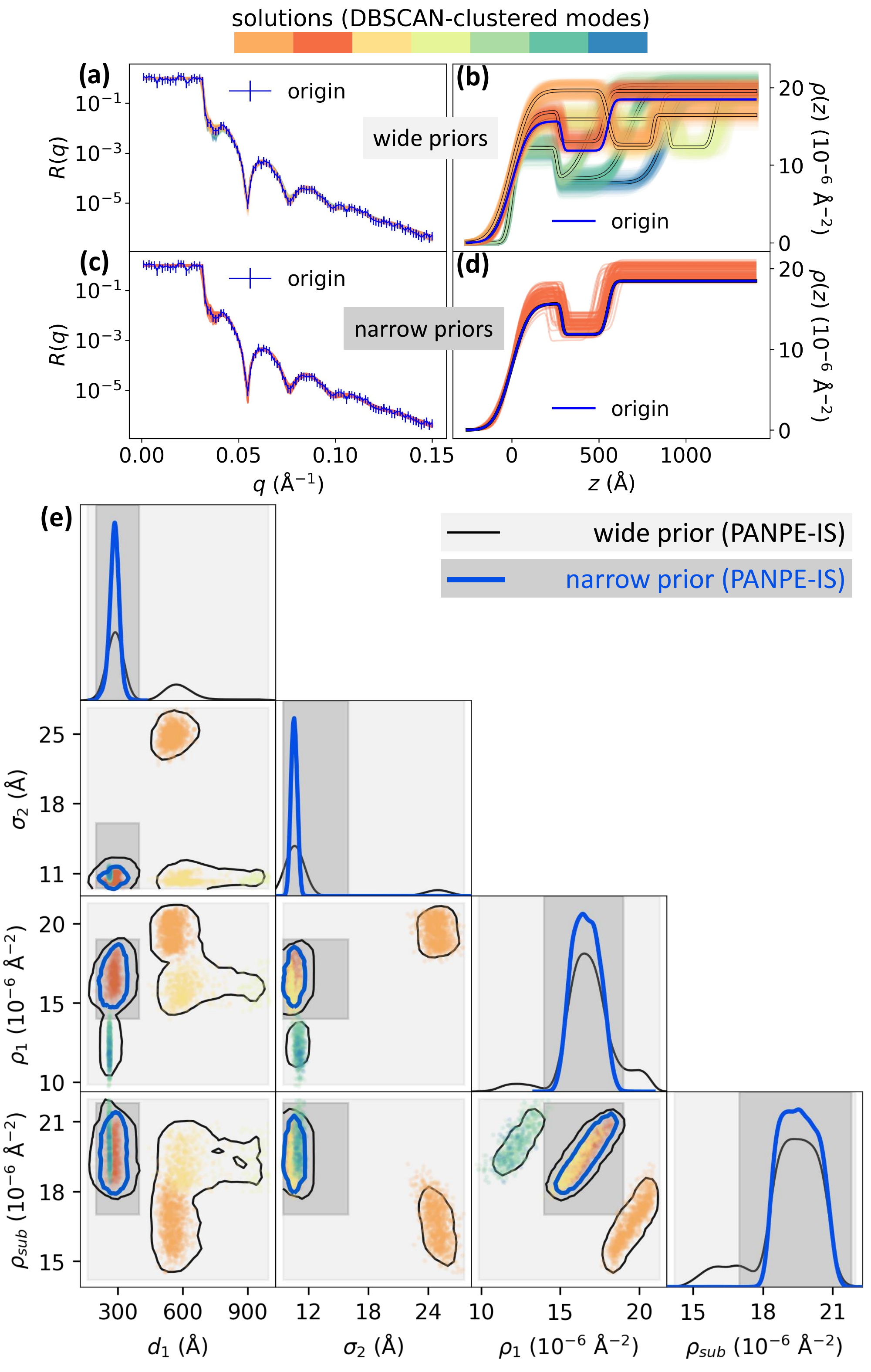}
            \caption{Multimodal posterior distribution obtained by PANPE-IS on a simulated reflectivity curve with 10 free parameters for a two-layer structure. The neural network produces results in accordance with the provided prior information, identifying (a) multiple solutions for a ``wide'' prior distribution and (c) a single distributional mode for a ``narrow'' prior (gray dashed lines). Colors denote distinct distributional modes obtained by clustering samples. The corresponding reflectivity curves (b) and (d) enable real-time likelihood-based refinement, resulting in accurate posterior estimation. The corner plot (e) shows the resulting marginalized 4d distributions obtained for both priors along with the colored samples related to the colored profiles in (a).}
            \label{fig:fig2}
        \end{figure}

        \paragraph{Multiple modes in the posterior} Figure \ref{fig:fig2} showcases the inference results obtained with PANPE-IS for a curve simulated from a two-layer structure. For reflectometry, generated samples $\{\bm\theta_i\}_{i=1}^N$ represent SLD profiles that could potentially produce the measured data according to the neural network. Figure \ref{fig:fig2}(a-b) shows the reflectometry curve colored in blue on the left and $N = 1000$ colored SLD profiles obtained from our model on the right with a wide prior distribution. The colors indicate 7 distinct solutions (distributional modes) separated via DBSCAN clustering for better visualization. The blue SLD profile represents the ``true'' structure used to simulate the reflectometry curve. The forward reflectometry simulations enable immediate visual validation of the result. In Figure \ref{fig:fig2}(a) the studied blue reflectometry curve is superimposed with the colored curves simulated from the NN-produced SLD profiles. The colored reflectometry curves are mostly invisible since they overlap very well with the original curve and each other, despite a diverse variety of corresponding SLD profiles. Figure \ref{fig:fig2}(e) provides an alternative visualization of samples and the estimated posterior on a corner plot. For visual purposes, only 4 out of 10 parameters are shown here. 
        
        \paragraph{Efficiency gain from prior amortization} Figure \ref{fig:fig2}(c-d) shows the inference result for the same reflectivity curve, but with narrower prior distribution, resulting in a single mode. In this case, the prior amortization allows excluding all the samples outside the domain of a more informative prior distribution. 

        The necessity for prior amortization might not be immediately obvious when likelihood evaluation is fast. A viable alternative could seem to be training a standard NPE model across a wide parameter range without prior amortization and refining results via likelihood-based methods later. For uniform priors, calculations of likelihood are not even required: samples outside the prior domain can be simply rejected without the need of likelihood evaluation. However, the main issue with rejection sampling is its low acceptance rate. In practice, this quantity can be exceedingly small. In our case, an acceptance rate of less than one in a million applies to about 70 \% of the synthetic test data. Consequently, in order to obtain a single sample within the prior domain, an immense number of samples would need to be generated through neural network evaluations, making this approach essentially inapplicable. We also illustrate this problem on an experimental data in Section \ref{sec:results:exp-xrr}.

        \begin{figure}
            \centering
            \includegraphics[width=0.6\linewidth]{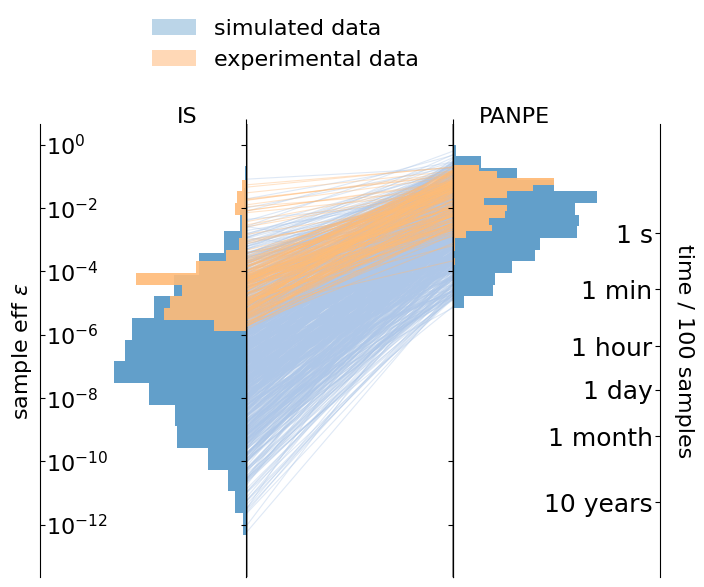}
            \caption{Sample efficiencies for conventional importance sampling (left) and our PANPE-IS model (right) on a test dataset of 1000 simulated curves (blue) and a experimental dataset of 208 X-ray reflectometry curves  (orange). An additional axis on the right-hand side indicates the estimated time it takes to generate 100 effective samples (ESS) on our hardware with the efficient GPU-accelerated reflectometry simulator (see text). Both the simulated and experimental data consist of two-layered structures with 10 parameters, but in the experimental data, only the top layer is unknown, as the parameters of the silicon substrate and the silicon oxide layer are largely constrained through their respective priors.}
            \label{fig:performace-parallel-plot}
        \end{figure}

        \paragraph{Sample efficiencies on the simulated dataset} We evaluate PANPE-IS on a set of 1000 simulated test samples. These samples are generated following the same procedure as outlined for the training data in Methods \ref{sec:methods:training}. Each curve has different $q$ discretization and is accompanied by its own prior distribution $p(\theta \newmid \bm\phi)$. 
        
        We assess the performance of the model on each test sample by evaluating its sample efficiency $\eeff$:

        \begin{equation}\label{eq:eff-weights-main}
            \eeff = \frac{\left(\overline{w_i}\right)^2}{\overline{\left(w_i^2\right)}} \text{ ,}  \quad w_i = \frac{p(\bm{R} \newmid \bm\theta_i)p(\bm\theta_i \newmid \bm\phi)}{p_{\mathrm{NN}}(\bm\theta_i \newmid \bm{R}, \bm\phi)} \text{ ,} \quad \bm\theta_i \sim p_{\mathrm{NN}}(\bm\theta \newmid \bm{R}, \bm\phi) \text{ ,}
        \end{equation}
        where $w_i$ are the importance weights. In practice, during inference, importance weights can be employed for Monte Carlo estimates $\mexpect_{\bm\theta \sim p(\bm\theta \newmid \bm{R}, \bm\phi)}[f(\bm\theta)] \approx \left(\sum_{i=1}^N f(\bm\theta_i) w_i \right)/\left(\sum_{i=1}^N w_i\right)$. The sample efficiency $\eeff$ effectively quantifies the efficient sample size $\mathrm{ESS} = N \cdot \eeff$ as a share of the total number of samples and, therefore, determines the time required for obtaining a desired ESS. By using our efficient reflectometry simulator, we are able to obtain $\mathrm{ESS} = 100$ for $\eeff= 10^{-4}$ for less than a minute, but the same $\mathrm{ESS}$ would take more than a month for $\eeff = 10^{-10}$.
        
        Each test sample is characterized by its prior distribution, which results in different complexity of the inference task: wider prior distributions that simulate the cases of higher uncertainty about the studied structure are generally more complex to analyse using conventional methods. We quantify this complexity through the sample efficiency of the conventional importance sampling (IS) method with prior acting as a proposal distribution. In this case, it replaces the ``neural'' proposal distribution in the Equation \ref{eq:eff-weights-main}, and the respective importance weights simply equal to the likelihood $w^{\mathrm{IS}}_i = p(\bm{R} \newmid \bm\theta_i)$, where $\bm\theta_i \sim p(\bm\theta | \bm\phi)$. We discuss how we estimate low sample efficiencies for IS in Methods \ref{sec:methods:low-sample-efficiency}.

        Figure \ref{fig:performace-parallel-plot} shows sample efficiency distribution of conventional IS on the left-hand side and our PANPE-IS on the right hand side, with lines in the middle connecting the same test samples and indicating the difference in sample efficiency between the two methods. Blue color corresponds to the synthetic test dataset (we discuss the experimental data in the next Section). The axis on the right shows an average time required to obtain $\mathrm{ESS} = 100$ for different sample efficiencies on a single graphics card, NVIDIA GeForce RTX 2080 Ti, with our GPU-accelerated reflectometry simulator. It shows that our PANPE model can perform inference in under a minute where the conventional IS approach would require days and even months of computation.

        \paragraph{Evaluating PANPE performance without refinement} Additionally, we evaluate the performance of raw PANPE estimates without likelihood-based refinement. Specifically, we evaluate the quality of marginal distributions, i.e. one-dimensional distributions $p_{\mathrm{NN}}(\theta_j | \bm{R}, \bm\phi)$ for each $j$th parameter. For that, we perform standard Kolmogorov-Smirnov tests that employ the ``true'' parameters $\bm\theta$ used for simulating the test data \cite{dax2021real}. These tests determine whether the true parameters could realistically be sampled from PANPE-generated marginal distributions by checking if their percentile scores are uniformly distributed. SM Figure 1 shows the p-p plots, and the obtained p-values demonstrate satisfactory performance on simulated data. These findings imply that one can rely on raw PANPE estimates derived from marginal distributions like means and variances for analysis. However, in the context of reflectometry, we always apply likelihood-based refinement as it is cost-effective, and importantly, enhances the accuracy of our estimates while also providing a means to evaluate their quality.

\subsection{Performance on experimental XRR data}\label{sec:results:exp-xrr}

    In this Section, we evaluate the performance of PANPE-IS on the largest publicly available reflectometry dataset \cite{pithan2022reflectometry_data} that has been previously employed for evaluating the performance of ML-based regression models \cite{Greco2019Fast, Greco2021Neural, Greco2022Neural}. 

    \paragraph{Experimental data} As experimental XRR data, we utilize 208 curves, accompanied by a manual analysis using a conventional fitting procedure via maximum likelihood estimation. This data originates from three online \textit{in situ} experiments conducted at different synchrotron facilities. Each experiment recorded in real time a process of growing an organic layer, specifically diindenoperylene (DIP), on a silicon substrate. DIP, an organic semiconductor, has gained interest due to its prospective uses in the field of electronics and photovoltaics \cite{Wagner2010}. Real-time XRR measurements can provide insights into growth processes of such thin films. Furthermore, this type of experiment can indeed benefit from rapid analysis. In this way, a machine learning-based solution was recently deployed for the first closed-loop XRR experiment \cite{Pithan2023loop}. However, the ambiguity problem presented limitations to the regression-based model, rendering our probabilistic method a potential successor in such closed-loop systems (see SM Figure 6).

    \begin{figure}[h]
        \includegraphics[width=0.9\linewidth]{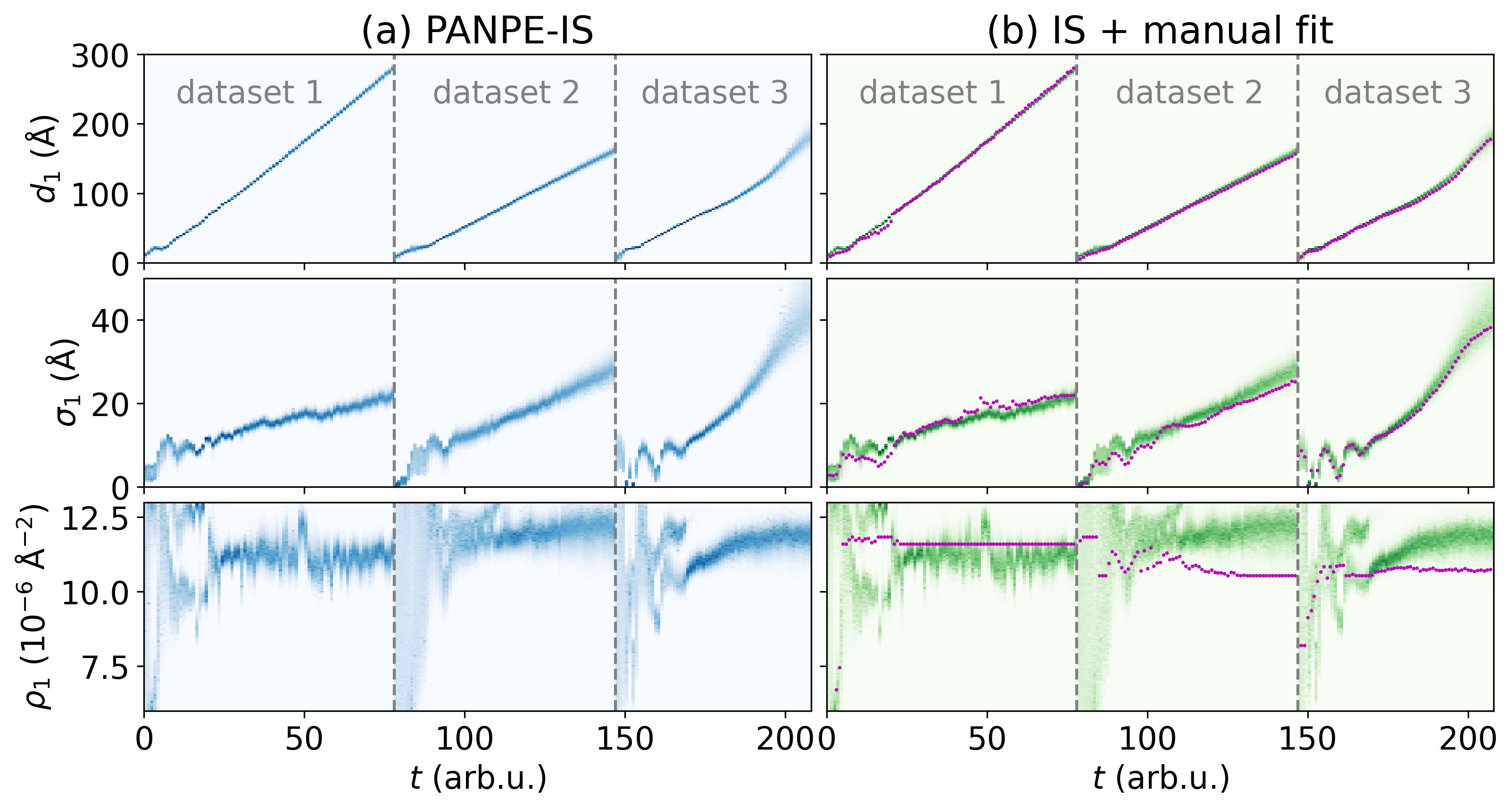}
        \caption{Marginal distributions of the thickness $d_1$, roughness $\sigma_1$, and density $\rho_1$ of the diindenoperylene (DIP) layer growing on a silicon substrate for three \textit{in situ} experimental XRR datasets obtained by our model (on the left) and via conventional importance sampling from prior distribution (on the right). The colors designate normalized probability densities. Purple dots correspond to conventional manual fits performed via differential evolution reported in \cite{pithan2022reflectometry_data}. SM Figure 4 shows time-dependent sample efficiencies and the log evidence estimations for both methods.}
        \label{fig:exp-marginals}
    \end{figure}

        \begin{figure}[H]\centering
            \includegraphics[width=\linewidth]{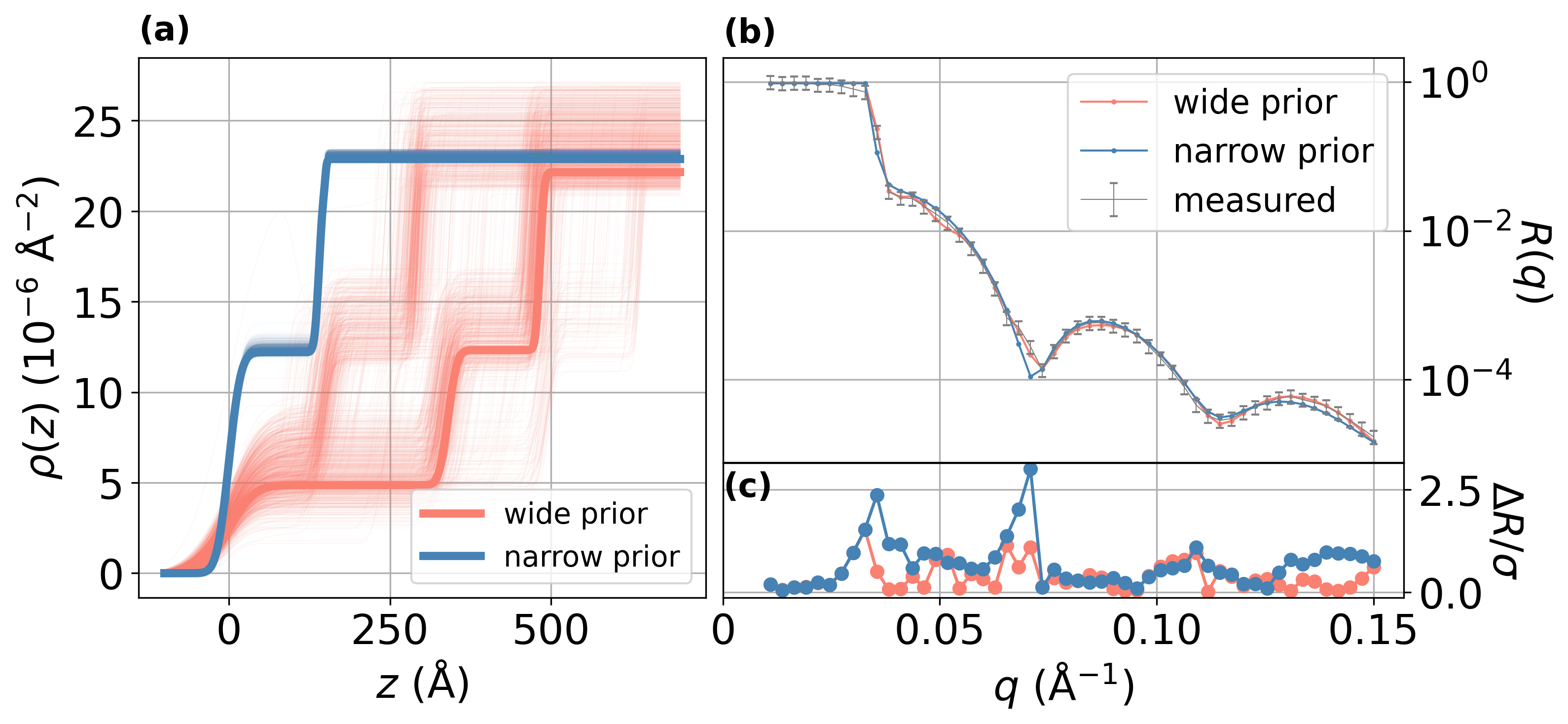}
            \caption{Experimental XRR curve analyzed using both a wide prior distribution that encompasses the entire training range (shown in red) and a narrow, physics-informed prior distribution (shown in blue). (a) SLD profiles associated with the PANPE-IS samples. Profiles with the highest likelihood are highlighted with bold lines. (b) The observed reflectivity curve (in gray) is compared with simulated curves that correspond to the maximum likelihood from both the narrow and wide prior distributions. While both fits are satisfactory, the unphysical solution (in red) has a likelihood that is more than $10^6$ times greater than its physical counterpart due to larger residuals (c). In this case, when trained solely with a wide prior distribution, the NPE network mainly samples unphysical solutions, which appear much more probable without the physics-informed prior. The use of prior amortization addresses this issue. }
            \label{fig:wide_narrow_priors_exp_curve}
        \end{figure}

    For each experimental curve, we set uniform priors based on a physical understanding of the experiment, aligning with conventional analysis. The parameters of the known silicon substrate and the oxide layer are essentially fixed by designating narrow ranges. In contrast, the parameters for the thickness $d_1$, roughness $\sigma_1$, and density $\rho_1$ of the growing organic layer have broader prior ranges due to uncertainty. Furthermore, as the film thickness $d_1$ increases, its ranges are set to increase linearly with time, in line with the anticipated growth rate. Although the physical model contains 10 parameters, in this case, the physics-informed prior information about the structure allows us to effectively constrain most of the parameters. Nonetheless, prior amortization allows us to use the same PANPE model that was applied to simulated data featuring two-layer structures.

    We also note that the datasets under consideration feature different $q$ ranges and resolutions. Nevertheless, a single model can process them due to the prior amortization that exploits the scaling invariance of the reflectometry data, as well as due to the amortized discretization of our model.
    
    \paragraph{Comparison with conventional data analysis} The defined prior distributions are narrow enough to enable conventional importance sampling for validation purposes. Figure \ref{fig:exp-marginals} displays the time-dependent marginal distributions for three parameters obtained by both our PANPE-IS model (on the left) and the conventional importance sampling (IS) method (on the right), showing equivalent solutions. SM Figure 2 demonstrates a corner plot with posterior estimates obtained for an experimental curve using PANPE, PANPE-IS, and PANPE-MCMC, where two refinement methods lead to equivalent distributions. 
    
    Despite the relative simplicity of the dataset due to a small number of free parameters, the resulting distributions feature two solution branches for the density parameter $\rho_1$ visible in Figure \ref{fig:exp-marginals} (see also SM Figure 3). The upper branches vanish beyond a certain time for each dataset, suggesting that the correct solution corresponds to the lower branches. Conventional fits, indicated by purple dots in Figure \ref{fig:exp-marginals}(b), deviate from the maximum likelihood, underscoring the relevance of probabilistic methods even in such straightforward cases. 
    
    Figure \ref{fig:performace-parallel-plot} (orange color) demonstrates sample efficiencies on the experimental data for the conventional IS and our PANPE solution. Notably, in the case where most parameters are effectively known and constrained, conventional IS can be a practical solution, unless a real-time analysis is required. For most of the considered experimental data, our model performs the analysis in under a second, where IS may require tens of minutes per sample. Several curves where IS is almost as efficient as PANPE-IS correspond to the beginning of the growth process, when there is essentially still no organic layer, and the other parameters are known. Therefore, the axis for IS sample efficiency in Figure \ref{fig:performace-parallel-plot} can be approximately divided into ranges: $\eeff^{\mathrm{IS}} > 10^{-3}$ for ``zero-layer'' structures, $\eeff^{\mathrm{IS}} \in [10^{-6}, 10^{-3}]$ for ``one-layer'' structures that constitute the rest of the experimental curves, and $\eeff^{\mathrm{IS}} < 10^{-6}$ for more complex, simulated two-layer structures. The respective estimated inference time axis in Figure \ref{fig:performace-parallel-plot} suggests that pure conventional likelihood-based methods become largely impractical for two-layer structures. On the other hand, PANPE-IS delivers accurate and reliable solutions in under a minute for most of these cases.

    SM Figure 5 displays the inference results for a simulated four-layer structure with 16 parameters, revealing a highly ambiguous outcome (the result is obtained via PANPE-IS using an additional model trained on four-layer structures). This example underscores the increasing complexity and ambiguity in reflectometry analysis as the number of free parameters grows. To maintain the high sample efficiency of PANPE in these more challenging scenarios, it is necessary to enhance the capacity of the density estimator. This enhancement could be achieved not only by enlarging the size of the neural network, but also through the use of more sophisticated density estimators, such as continuous normalizing flows \cite{NEURIPS2018_69386f6b, lipman2022flow, wildberger2024flow}, without necessitating significant changes to the overall framework presented in this paper. Improving performance for more complex scenarios may also involve customizing the hyperprior distribution and other solution aspects, such as $q$ discretization, to better suit the specific application and experimental conditions.
    
    \paragraph{Role of physics-informed priors} Figure \ref{fig:wide_narrow_priors_exp_curve} illustrates the essential role of prior amortization in the analysis of the presented experimental data. Without providing the physics-informed prior distribution to the neural network, the resulting samples span all possible solutions (represented by the red SLD profiles in Figure \ref{fig:wide_narrow_priors_exp_curve}a) within the expansive prior distribution that covers the complete training range. Yet, all of these solutions are unphysical, due to factors such as a too thick oxide layer. The physical solution, depicted by the blue SLD profiles, is practically unattainable without prior amortization given that the share of samples within the corresponding narrow prior is less than $10^{-6}$. This scenario emphasizes once more the essential role of incorporating additional physical information into inverse scattering problems with missing phase.

    \paragraph{Adaptive $q$ discretization} The ability of our model to support arbitrary $q$ discretization significantly broadens its applicability. In this way, the number of $q$ points in the analyzed X-ray data ranges from 25 to 52, yet it is processed by the same model. We note that an alternative approach involving interpolation to conform to a fixed $q$ axis can generally lead to missed solutions. For instance, if a model trained on 52 $q$ points is subsequently tested on experimental data comprising only 25 $q$ points, interpolating these data to 52 points could create a falsely narrow distribution. This occurs because the interpolation process artificially adds ``information'' that the original experimental data does not possess, compromising the guarantee of the coverage property.

    SM Figure 6 demonstrates another relevant experimental scenario that requires adaptive $q$ discretization: while performing an X-ray reflectometry measurement by sequentially measuring intensities at distinct points, one can use the model to analyze the data at any particular moment. This analysis can inform whether additional data are needed to reduce uncertainty and can even guide the selection of the next most informative $q$ point to measure. Obtaining such points is straightforward using reflectometry curves simulated using parameters sampled via PANPE-IS. 

    \subsection{Combination of multiple neutron reflectometry measurements}\label{sec:results:neutron}

    In this Section, we demonstrate that PANPE can be successfully employed for the simultaneous analysis of several combined neutron reflectometry datasets, which is an indispensable tool in the context of contrast variation (e.g. using different levels of deuteration \cite{Heinrich2009, Cai2024}).

    \begin{figure}[h]
        \includegraphics[width=\linewidth]{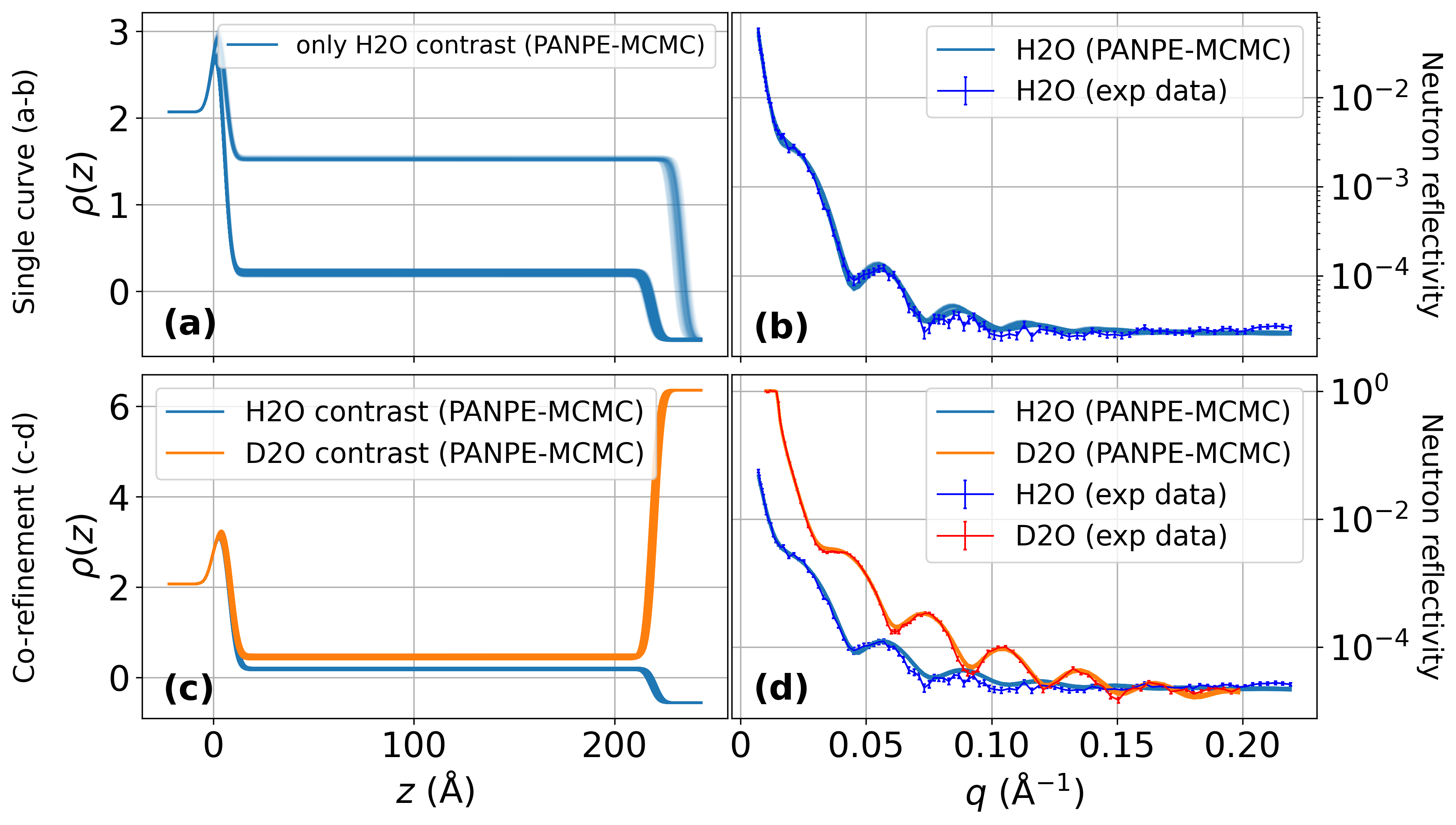}
        \caption{Co-refinement of two neutron reflectometry measurements of a polymer on a silicon substrate using H$_2$O and D$_2$O contrasts, conducted via PANPE-MCMC. (a-b) An analysis of a single measurement with H$_2$O contrast (b) yields two distinct solutions and their corresponding SLD profiles (a). (c-d) Co-refinement of the measurements using H$_2$O contrast (blue) and D$_2$O contrast (orange) resolves ambiguity by eliminating one of the solutions. Data and priors are sourced from the \textit{refnx} package \cite{Nelson2019refnx}. The combination of several measurements is enabled in this case by the use of parameter-conditioned posterior estimation.}
        \label{fig:neutron-data-corefinement}
    \end{figure}

    \paragraph{Co-refinement of neutron data} A common method to resolve ambiguity involves combining measurements taken under controlled variation of experimental conditions. In the context of reflectometry, such conditions can be determined by the different energies of the X-ray beam (i.e. anomalous scattering near an absorption edge), polarizations of neutrons for magnetic materials, or by employing different contrasts via changing materials adjacent to the sample. 

    We demonstrate this co-refinement procedure using publicly available neutron data \cite{Nelson2019refnx}. Specifically, we consider two neutron reflectometry measurements of a polymer on a silicon substrate performed separately with H$_2$O ($\rho = -0.56 \cdot 10^{-6} \ \angstrom^{-2}$) and D$_2$O ($\rho = 6.36 \cdot 10^{-6} \ \angstrom^{-2}$) solvents. Neutron reflectometry differs from X-ray data in several aspects, such as instrumental resolution, high scattering background, and negative SLD. Therefore, we trained an additional PANPE model for neutron data that incorporates these features and has 11 free parameters that now also include scattering background. 

    \paragraph{Constructing proposal distribution from several measurements} The likelihood for two measurements $\bm{R}_{\mathrm{H_2 O}}$ and $\bm{R}_{\mathrm{D_2 O}}$ is a product $p(\bm{R}_{\mathrm{H_2 O}} \newmid \bm\theta)p(\bm{R}_{\mathrm{D_2 O}} \newmid \bm\theta)$. The corresponding posterior distribution cannot be directly estimated by (PA)NPE model unless specifically trained on such combined measurements. However, using our model trained on single curves, one can combine two (or more) sets of samples generated independently for each measurement for constructing a proposal distribution $\frac12 (p_{\mathrm{NN}}(\bm\theta \newmid \bm{R}_{\mathrm{H_2 O}}) + p_{\mathrm{NN}}(\bm\theta \newmid \bm{R}_{\mathrm{D_2 O}}))$. Such a proposal distribution exhibits the probability mass-coverage property and can be further refined via likelihood-based methods for obtaining a reliable and accurate posterior distribution. 

    However, additional complications arise when only a subset of the estimated parameters -- namely, the unchanged parameters of the studied sample, $\bm\theta^{\mathrm{shared}}$ -- are shared among multiple measurements. Other parameters, $\bm\theta^{\mathrm{unique}}$, such as background, misalignment, and different contrast densities, are unique to each measurement. Thus, the estimated parameters are expressed as $\bm\theta = [\bm\theta^{\mathrm{shared}}, \bm\theta^{\mathrm{unique}}_{\mathrm{H_2 O}}, \bm\theta^{\mathrm{unique}}_{\mathrm{D_2 O}}]$. As a result, a subset of parameters generated by the model for the first measurement, $[\bm\theta^{\mathrm{shared}}, \bm\theta^{\mathrm{unique}}_{\mathrm{H_2 O}}] \sim p(\bm\theta \newmid \bm{R}_{\mathrm{H_2 O}})$, is incomplete as it lacks the subset of parameters unique to the second (other) measurement(s) $\bm\theta^{\mathrm{unique}}_{\mathrm{D_2 O}}$, and vise versa. The solution involves sampling the remaining parameters from the \textit{parameter-conditioned} posterior distribution: $\bm\theta^{\mathrm{unique}}_{\mathrm{D_2 O}} \sim p(\bm\theta \newmid \bm{R}_{\mathrm{D_2 O}}, \bm\theta^{\mathrm{shared}})$. This conditional distribution is not provided by NPE and typically necessitates training additional models. 
    
    The reparameterization operation that we introduce as a part of our prior-amortized approach offers a means to approximate such conditional probability densities with the same model by setting very narrow priors, essentially fixing the required parameters. However, this approach provides only samples and not density evaluation required for importance sampling refinement (see Methods \ref{sec:methods:panpe}). Therefore, we use PANPE-MCMC for likelihood refinement of the combined posterior in this case.

    \paragraph{Inference results for combined measurements}     Figure \ref{fig:neutron-data-corefinement} demonstrates the results of the PANPE-MCMC analysis of a single neutron reflectivity curve measured with H$_2$O contrast (a-b), as well as the joint analysis of two measurements incorporating both H$_2$O and D$_2$O contrasts (c-d). The single measurement yields two distinct solutions (Figure \ref{fig:neutron-data-corefinement}a), one of which is (implicitly) ruled out when performing a co-refinement of two measurements (Figure \ref{fig:neutron-data-corefinement}c), thereby resolving ambiguity in data interpretation. Non-zero ambient density is processed as discussed in Methods \ref{sec:methods:invariance}. 
    
    In the case of the neutron data analyzed in this work, the parameters unique for each measurement $\bm\theta^{\mathrm{unique}}$ include scaling misalignment, background, and densities of contrasts. The shared parameters correspond to the constant parameters (i.e. those that do not change in-between these measurements) of the system under study, such as thicknesses, roughnesses, and densities of Si and SiO$_2$. Notably, following the parameterization described in the \textit{refnx} package, we account for the volume fraction $v_{\mathrm{solv}} \in (0, 1)$ of the solvent that modifies the SLD of the polymer layer according to $\rho = (1 - v_{\mathrm{solv}})\rho_{\mathrm{polymer}} + v_{\mathrm{solv}}\rho_{solv}$. We regard these densities $\rho$ as parameters unique to each measurement, utilizing them to compute $v_{\mathrm{solv}}$ and $\rho_{\mathrm{polymer}}$. The mixture of solvent with polymer results in a small difference of polymer SLDs for measurements with different contrasts in Figure \ref{fig:neutron-data-corefinement}.
    
    It is worth noting that in certain cases of application-specific parameterization on an SLD profile, it might be more practical to retrain the model using a more suitable parameterization, but it is not necessary in this case.  In general, the demonstrated approach can be equally well applied to other cases of parameter co-refinement such as polarized neutron reflectometry, XRR measurements with different energies, or other similar applications that require combining several measurements.

    \section{Discussion}

        In reconstructing physical systems from scattered intensities, the phase problem poses a fundamental challenge. This problem is encountered in numerous scattering techniques, including X-ray and neutron reflectometry. The prevailing standard in reflectometry analysis is maximum likelihood estimation, which uses a differential evolution-based search over the parameter space. This method, by design, produces a single system reconstruction, even when multiple solutions exist due to phase loss, making it inherently unreliable. In contrast, Bayesian inference provides a foundational pathway to a more reliable analysis by inherently accounting for all possible solutions as a distribution over the considered physical parameters. However, despite its conceptual advantages, conventional likelihood-based Bayesian methods struggle with high-dimensional parameter spaces and multimodality, often falling short in real-world experimental contexts. This underscores the pressing need for more efficient and reliable Bayesian methods in reflectometry analysis.
        
        In this work, we present a new approach that enables reliable, accurate, and fast Bayesian reflectometry analysis. The high inference speed essential for various experimental contexts is achieved via neural network-based amortization. The training strategy reliably provides an approximate posterior distribution that fully covers the true posterior, while likelihood-based inference is subsequently employed to obtain the accurate distribution. 
        
        Importantly, our method enables reliable reflectometry experiments: it can guide experimentalists by indicating whether the remaining ambiguity requires additional measurements, thereby directing the experiment until the singular physical solution is determined. Furthermore, preparing experiments in advance by investigating ambiguities in simulated settings becomes possible. 
        
        Prior amortization crucially broadens the applicability of NPE to multiple experimental settings. In the context of reflectometry, constraining the parameter space based on individual characteristics of a studied structure is essential to resolve ambiguity. As shown, prior amortization allows to infer a physical solution that can feature a million times smaller likelihood and filter out an unphysical solution that is yet legitimate in other experimental settings or for other systems. This, in particular, underscores the importance of prior amortization when applying simulation-based inference methods to scattering problems with phase loss. Complex parameterizations of prior distributions can be used  and should be investigated in future research aiming at adapting PANPE to specific applications.  

        Real-world benchmarks are valuable for the developing field of simulation-based inference. In this context, reflectometry analysis offers a notable benchmark, characterized by challenging multimodal distributions and bolstered by an efficient simulator. Importantly, it can be straightforwardly scaled up by increasing the number of layers in the physical model. Simulation-based inference is often seen as advantageous for applications where likelihood is costly or intractable to evaluate. Reflectometry does not fit into this category and it exemplifies the broader utility of SBI methods beyond intractable likelihoods due to the coverage property and acceleration of inference through amortization. Moreover, the combination of SBI with likelihood-based methods presents as the optimal way to both preserve the coverage and achieve high accuracy.
        
        Prior amortization can be beneficial for multiple applications, especially in experimental science like scattering where various experimental scenarios require adaptive prior distributions. In this manner, our method is suitable for a wide range of scientific experiments that permit simulation-based inference.

\section{Methods}

\subsection{Parameterization of the SLD profile}\label{sec:methods:theta}
    
    We consider the standard parameterization of the SLD profile of a layered structure with $n_l$ layers through parameters $\bm\theta = \{\bm{d}, \bm{\rho}, \bm\sigma, \Delta R, \Delta q, \log_{10} (R_0)\}$, and we primarily consider two-layer structures $n_l = 2$. Here $\bm{d} = \{d_1, d_2\}$ are layer thicknesses in the top-bottom order, $\bm{\rho} = \{\rho_1, \rho_2, \rho_{\mathrm{sub}}\}$ are densities of two layers and the substrate, and $\bm{\sigma} = \{\sigma_1, \sigma_2, \sigma_{\mathrm{sub}}\}$ are roughnesses of three interfaces modeled via Névot–Croce factors. We exclude absorption in this work due to our focus on organic materials, but it can be straightforwardly included into our framework. Additionally, we consider standard misalignment parameters: normalization misalignment $\Delta R$ and systematic misalignent of the $q$ axis $\Delta q$. We only consider the parameter for scattering background $\log_{10} (R_0)$ for neutron data (see Section \ref{sec:results:neutron}), which results in 11 parameters. Consequently, the model for X-ray data has 10 free parameters.

\subsection{Equivariant transformations in reflectometry}\label{sec:methods:invariance}

        Reflectometry features several equivariant transformations that can be considered to improve the performance of an amortized machine learning solution. In this section, we discuss these transformations and how we incorporate them into our PANPE model.

        \paragraph{Unit-based scaling equivariance} Reflectometry simulation $R(q, \bm\theta) = R(q , \bm{d}, \bm\sigma, \bm\rho)$ features an invariant scaling transformation that represents the change in employed parameter units $u$:

        \begin{equation}\label{eq:refl-invariant-transform}
            T_u(R(q, \bm{d}, \bm\sigma, \bm\rho)) = R(q, \bm{d}, \bm\sigma, \bm\rho) \text{, }
        \end{equation}
        where 
        \begin{equation}
            T_u(R(q, \bm{d}, \bm\sigma, \bm\rho)) \equiv R(q\cdot u, \bm{d} / u, \bm\sigma / u, \bm\rho \cdot u^2) \text{,}
        \end{equation}
        and $u \in \mathbb{R}_{> 0}$ is a positive value that defines the employed units. The standard units are inverse Angstroms ($\angstrom^{-1}$) for $q$ values, Angstroms ($\angstrom$) for layer thicknesses $\bm{d}$ and roughnesses $\bm\sigma$, and inverse squared Angstroms ($\angstrom^{-2}$) for (scattering length) densities $\bm\rho$. If we set $u = 1$ for these standard units, for instance, the transformation with $u = 10$ would correspond to the change of units from Angstroms to nanometers, which does not alter the resulting reflectivity curve. Similarly, $u=2$ doubles the $q$ range, halves thicknesses and roughnesses, and increases densities by a factor of $u^2 = 4$, leaving the reflectivity curve unchanged. 

        This invariance of the reflectometry simulator leads to the \textit{equivariance} of the density estimator under the joint unit transformation of input and parameters. Specifically, stretching or squeezing the $q$ axis in the input data \textit{and adjusting the input prior parameters} $\bm\phi$ accordingly should result in respective transformations of the parameters $\bm\theta$ as per \autoref{eq:refl-invariant-transform}. We note that the transformation of prior distribution is commonly required in order to preserve equivariance in the density estimator. Applying this transformation indeed requires prior amortization.

        To incorporate this equivariance into our model, we can \textit{standardize the ``pose'' of the data} \cite{Dax2022GNPE} (the terminology is adopted from computer vision) to simplify the problem for the neural network. We do so by fixing the $q$ range, on which our model is trained. During inference, we first preprocess the data by applying the transformation from \autoref{eq:refl-invariant-transform} so that the measured $q$ range matches the standard one. The corresponding scaling factor is the ratio of two ranges $u = q_{\mathrm{max}} / q_{\mathrm{exp}}$. We use this scaling factor to apply the respective transformation on the prior parameters $\bm\phi$. After obtaining samples from the PANPE model, we rescale the parameters back using $u^{-1}$. Respectively, the probability densities are corrected by a constant Jacobian determinant of the transformation, which equals $u^{-2}$ in our case. 

        We note that this property should also be taken into account when considering parameter ranges for training. For instance, some unreasonably large parameter ranges that might seem unphysical, such as density values that do not correspond to any known materials, can be practically justified since they correspond to smaller densities when scaling the $q$ axis. This relation is illustrated in SM Figure 8.

        \paragraph{Density shifting equivariance} Reflectometry is sensitive to density contrasts rather than absolute density values. As a result, shifting all the densities $\bm\rho$ (SLDs) in the system, including the ambient and the substrate, is an invariant operation that does not change the resulting reflectivity curve. In the context of the density estimator, this leads to the equivariant operation: shifting the respective prior parameters $\bm\phi$ should result in the shift of the parameters $\bm\theta$ (specifically, layer densities $\bm\rho$). 

        We employ this property by defining a natural standard pose in a form of the zero ambient density, on which the model is trained. During inference, the data with non-zero ambient density is first preprocessed by shifting the densities (i.e., the respective prior parameters) so that the ambient density becomes zero. We apply this transformation for neutron reflectometry in \autoref{sec:results:neutron}. The Jacobian determinant of this transformation equals to 1. 

        \paragraph{Misalignment shifting equivariances} The misalignment parameter $\Delta R$ results from the incorrect normalization when calculating reflected intensities as per $R(q, \bm\theta) \cdot (1 + \Delta R)$, which effectively shifts the reflectometry curve in the ``vertical'' direction in the log space, resulting in an equivariant shifting transformation. A natural ``standard pose'' in this case corresponds to $\Delta R = 0$. We note that in this case the standard pose depends on the (unknown) parameter $\Delta R$ rather than the data and cannot be performed as a one-step preprocessing. This scenario is similar to the one considered previously \cite{Dax2022GNPE}, where an iterative inference scheme is proposed that allows converging to the standard pose. In our case, the range of the misalignment parameter is already very limited, making a direct application of the iterative scheme impractical. Our preliminary tests suggest that this does not lead to improved performance, so we do not utilize this equivariance in our solution. The same applies to the other misalignment parameter $\Delta q$.

\subsection{PANPE training}\label{sec:methods:panpe}

    We amortize Bayesian inference for a class of prior distributions $p(\bm\theta \newmid \bm\phi)$, parameterized by $\bm\phi$. To specify the range of priors for which the model is trained, we introduce the hyperprior distribution -- a distribution over prior parameters $p(\bm\phi)$, that generally depends on the range of anticipated applications and can reflect various physical and practical parameter constraints. We set it to cover a broad range of practical scenarios where some of the parameters are known better (with lower uncertainty) than others. 

    The training process of the PANPE model involves adjusting the trainable parameters, denoted as $\bm{w}$, of the neural network to minimize a forward Kullback–Leibler (KL) divergence between the true posterior distribution $p(\bm\theta \newmid \bm\phi)$ and the flow-based density estimator $p_{\bm{w}}(\bm\theta \newmid \bm\phi)$:

    \begin{equation}\label{eq:loss-kl}
         L(w) = \mexpect_{p(\bm\phi)}\mexpect_{p(\bm\theta \newmid \bm\phi)p(R \newmid \theta)} \big[\log \left(\frac{p(\bm\theta | R, \bm\phi)}{p_{w}(\bm\theta | R, \bm\phi)}\right)\big] = \mexpect_{p(\bm\phi)}[D_{KL}(p(\bm\theta | R, \bm\phi) || p_{w}(\bm\theta | R, \bm\phi))] \text{. }
    \end{equation}
    The loss in Equation \ref{eq:loss-kl} is approximated by the Monte Carlo estimation:
    \begin{equation}\label{eq:loss-mc}
         L(w) \approx \sum_i^N [- \log (p_{w}(\bm\theta_i | \bm{R}_i, \bm\phi_i))] + \text{const, }
    \end{equation}    
    where the constant does not depend on the model parameters $\bm{w}$. Evaluating \autoref{eq:loss-mc} requires evaluating log density $\log (p_{w}(\bm\theta_i | \bm{R}_i, \bm\phi_i))$ and drawing samples $\{\bm\phi_i, \bm\theta_i, \bm{R}_i \}_i^N$ from the training distribution $p(\bm\phi, \bm\theta, \bm{R}) \propto p(\bm{R} | \bm\theta)p(\bm\theta | \bm\phi)p( \bm\phi)$. 

    Exact density evaluation, and consequently the utilization of the forward KL divergence, is facilitated by normalizing flows, distinguishing them from many variational architectures. The forward KL divergence is a mass-covering loss, meaning that the optimized density density covers the whole support of the target distribution (otherwise the loss diverges), thereby ensuring no distributional modes are missed in the posterior estimation. Hence, although the training scheme of PANPE is independent of the specific architecture of the density estimator, the selection of normalizing flows as the density estimator and the corresponding loss function is crucial for the method's reliability. It is also noteworthy that some other recent density estimators exhibit the mass-covering property and can thus be integrated into the PANPE framework for future reflectometry applications.

    Sampling from the training distribution can be performed in two principled ways. The first one is employed in this paper and it goes as follows:

    \begin{enumerate}
        \item First, the prior parameters are sampled from the hyperprior distribution ($\bm\phi_i \sim p(\bm\phi)$) defining the corresponding prior $p(\bm\theta \newmid \bm\phi_i)$.
        \item Then, the parameter $\bm\theta_i$ is sampled from this specific prior distribution, $\bm\theta_i \sim p(\bm\theta \newmid \bm\phi_i)$.
        \item Finally, the corresponding reflectometry curve, $\bm{R}_i \sim p(\bm{R} \newmid \bm\theta_i)$, is sampled from the likelihood.
    \end{enumerate}
    
    We note that a similar training scheme, involving sampling from a hyperparameter distribution, has been previously employed in group-equivariant neural posterior estimation (GNPE) \cite{Dax2022GNPE}.

    The second possible way of sampling from the training distribution relies on the relation $p(\bm\phi)p(\bm\theta | \bm\phi) = p(\bm\phi | \bm\theta)p(\bm\theta)$. Therefore, instead of first sampling prior parameters $\bm\phi$ and then parameters $\bm\theta$, this order can be reversed to sample (potentially, multiple sets of) prior parameters $\bm\phi$ that correspond to the same parameters $\bm\theta$, hence the same simulations. The gain in this case comes from the opportunity to re-use the same simulations by coupling them with different priors and potentially reduce the required number of simulations, which is critical for certain applications. Since this is not applicable to reflectometry, we do not investigate this method any further. We only note that its implementation would involve additional Bayesian inference $p(\bm\phi | \bm\theta) \propto p(\bm\phi)p(\bm\theta | \bm\phi)$, the complexity of which depends on the chosen prior parameterization $p(\bm\theta | \bm\phi)$. For instance, in the case of the parameterization employed in our work, the inference can even be performed analytically (via inverse transform sampling).

    As an optional improvement of our method, we introduce a reparameterization transformation $\Tilde{\bm\theta} = T_{\bm\phi}(\bm\theta)$ of the parameters $\bm\theta$ to effectively ``rescale'' the parameters according to the respective prior. The reparameterization is chosen to ensure that the prior for the rescaled parameters $p(\Tilde{\bm\theta})$ does not depend on the prior parameters $\bm\phi$. The flow-based model then is trained to perform inference on these rescaled parameters:

    \begin{equation}
    \begin{split}\label{eq:bayes3}
        p_{\mathrm{NN}}(\Tilde{\bm\theta} | \bm{R}, \bm\phi) \approx p(\Tilde{\bm\theta} \newmid \bm{R}, \bm\phi) \propto p(R \newmid \Tilde{\bm\theta}, \bm\phi) p(\Tilde{\bm\theta})  \text{ ,  } \\p_{\mathrm{NN}}(\bm\theta | \bm{R}, \bm\phi) = p_{\mathrm{NN}}(\Tilde{\bm\theta} | \bm{R}, \bm\phi) |\det J_{T_{\bm\phi}}| \text{ .}
    \end{split}
    \end{equation}

    This reparameterization effectively re-frames the problem as a standard Neural Posterior Estimation, only now the likelihood depends on both the parameters $\Tilde{\bm\theta}$ and the prior parameters $\bm\phi$. By doing so, we can now apply narrow priors without running into numerical issues. This approach accelerates the training process and decreases the number of samples generated outside the prior support. 

    Importantly, when setting the prior width of some parameter $\theta_j$ to zero, we effectively fix it. The respective parameter $\Tilde{\theta_j}$ estimated by the model does not influence the likelihood and is essentially trained to match the reparameterized prior distribution (uniform in our case). Consequently, the connection between the reparameterized space $\Tilde{\bm\theta}$ and the parameter space $\bm\theta$ becomes surjective. To sample from the lower-dimensional distribution conditioned on the fixed parameter $\theta_j$, we need to marginalize over the parameter $\Tilde{\theta_j}$ in the reparameterized space. It is not directly possible to evaluate density of a marginalized distribution in normalizing flows. We note that in the ideal scenario when the parameter $\Tilde{\theta_j}$ is uniformly distributed, density evaluation becomes straightforward. Our tests suggest that marginal distributions $p(\Tilde{\theta_j})$ can deviate from a uniform distribution in practice. However, sampling from a distribution marginalized over $\Tilde{\theta_j}$ is straightforward as it simply requires omitting the marginalized parameters. In this way, our reparameterization scheme enables us to sample from the parameter-conditioned posterior estimation.

\subsection{Trained models and parameter ranges}\label{sec:methods:training}

    \paragraph{Trained models} We present the main results for X-ray (XRR) and neutron (NR) reflectometry. Due to certain differences in the underlying physics and subsequent differences in the simulator, we have trained two PANPE models: one for XRR and another for NR. The results on the simulated data are presented for the XRR model. Most properties are shared between these models, except for the ranges of density parameters used (the scattering length density (SLD) for neutrons can be negative), the instrumental resolution (more pronounced in NR), and the presence of strong constant background scattering. Although we focus on these two models in the paper, we also show some examples from other models, such as those with four-layer structures.
    
    \paragraph{Parameter ranges}
    The training parameters are constrained by the predefined ranges. In this paper, we employ the following ranges shared by all the layers. Densities range within $[0,\, 60 \cdot 10^{-6}\, \angstrom^{-2}]$ for the XRR model and $\rho \in [-20 \cdot 10^{-6}\, \ \angstrom^{-2},\, 60 \cdot 10^{-6}\, \ \angstrom^{-2}]$ for the NR model. Thicknesses and roughnesses range within $ [0,\, 500\, \angstrom]$ and $[0,\, 50\, \angstrom]$, respectively. Additionally, we limit the maximum roughness of the interface by the half thickness of the thickest adjacent layer. For the misalignment parameters, the ranges are $[-2 \cdot 10^{-3} \ \angstrom^{-1}, 2 \cdot 10^{-3} \ \angstrom^{-1}]$ for $\Delta q$ and $[-5 \%, 5 \%]$ for $\Delta I$. Additionally, for the NR model, we introduce the (log) background parameter $\log_{10}(R_0)$, $R_0 \in [10^{-9}, 10^{-4}]$, which is set to $10^{-10}$ for XRR. 
    
   We note that due to the equivariant transformations discussed in Methods \ref{sec:methods:invariance}, during inference the model can also operate outside these parameter ranges. This also means that one can set rather unphysical training ranges, such as large roughness or density, to cover certain realistic scenarios at different $q$ ranges and ambient densities. 
        
    \paragraph{Hyperprior distribution} During training, the parameters $\bm\phi = \{\theta^{\mathrm{min}}_j, \theta^{\mathrm{max}}_j\}_{j=1}^{n}$ are generated as follows. First, for each parameter $\theta_j$, the width of the uniform prior $\Delta\theta_j = \theta^{\mathrm{max}}_j - \theta^{\mathrm{min}}_j$ is sampled, which generally can range from 0 to the the total parameter ranges introduced above. In this paper, we use the weighted sum of the uniform and the truncated exponential distribution for sampling prior widths. The latter term is added to better represent narrow prior widths corresponding to higher certainty in the prior knowledge. Then, the ``center'' of a prior $c_j = (\theta^{\mathrm{max}}_j + \theta^{\mathrm{min}}_j)/2$ is sampled uniformly within the allowed range. The parameters $\bm\phi$ are then calculated from $\Delta\theta_j$ and $c_j$. Finally, the upper bounds for interface roughnesses are rescaled to not exceed half the maximum thickness of adjacent layers \cite{Greco2022Neural}. This overall sampling scheme effectively defines the hyperprior distribution $p(\bm\phi)$.

    The test simulated data was produced using the same sampling procedure as the training data. However, some of the curves (less than $5\%$) were manually excluded from the test dataset since they exhibited pathological properties such as nearly zero contrast between the layers. This scenario essentially reduces the number of physical layers in the studied structure and results in exact linear correlation between thicknesses of these layers. The proper way to reduce the number of layers in PANPE would be by setting the thicknesses of redundant layers to zero.

    \subsection{Data simulations}\label{sec:methods:data-sim}

    During training, we simulate reflectometry data using the training parameters $\bm\theta_i$ sampled as per Methods \ref{sec:methods:training}. Each reflectometry simulation $\bm{R}_i \sim p(\bm{R} | \bm\theta_i)$, $\bm{R} = \{q_p, R(q_p), s_p\}_{p=1}^{n_q}$ is performed in several steps discussed below.

    \paragraph{Q discretization} First, $q$ values are sampled uniformly from the range $q_p \sim U(0, q_{\max})$ to enable arbitrary discretization. As discussed in Methods \ref{sec:methods:invariance}, the $q$ range corresponding to the ``standard pose'' is set equal to $q_{\max} = 0.15\, \angstrom^{-1}$. However, we can vary this during inference by exploiting the unit-scaling equivariant transformation. The number of points $n_q$ is also sampled uniformly $n_q \sim U(20, 64)$. In practice, it is implemented by masking out some of the input data from the model during training.
    
    Amortized discretization is generally necessary because the posterior can be highly sensitive to it in reflectometry applications. Our tests show that by fixing the $q$ discretization during training, we are able to considerably improve the model performance on the simulated data. However, different discretization of the experimental data necessitates interpolation procedure, that can deteriorate the performance of the model and generally lifts the mass-probability coverage guarantees, especially for experimental data with the lower number of points.

    Furthermore, amortized discretization is especially important in online XRR experiments, where time limitations constrain the number of measured $q$ points. It in principle enables closed-loop AI-guided measurements that enable choosing the most informative $q$ point to measure given the current data to speed up the overall process and be able to real-time phenomena with higher time resolution. 
    
    Nevertheless, fixing discretization might be beneficial for applications with more standardized experimental setup. Furthermore, we acknowledge possible ways to improve $q$ simulations for neutron data to better reflect the physical nature of the process and possibly even tailor it for the use at certain neutron sources.

    \paragraph{Measurement uncertainty} Next, we generate relative measurement uncertainties $s_p \sim U(5\%, 30\%)$, independently for each $q$ point. We treat these uncertainties as error bars that correspond to standard errors typically employed in reflectometry analysis. As a noise model, we employ the normal distribution as a common approximation of Poisson counting statistics for a high number of counts. We note that generally the use of the Poisson likelihood should be preferred in the case of low counts, which are especially frequent in neutron reflectometry. However, that requires reporting raw intensities, which are typically not included in the published data. 

    \paragraph{Reflectivity curves} Finally, we simulate reflectivity curves, using the generated parameters $\bm\theta = \{\bm{d}, \bm\sigma, \bm\rho, \Delta q, \Delta R, \log_{10}(R_0)\}$, $q$ points and measurement uncertainties $s$. We simulate curves in mini-batches using our parallelized GPU-accelerated PyTorch implementation of Abel{\`{e}}s transfer-matrix method \cite{Abeles1950La}:

    \begin{equation}
        R_p = \left(R(q + \Delta q, \bm{d}, \bm\sigma, \bm\rho) \cdot (1 + \Delta R) + R_0\right) \cdot e_p \text{, }
    \end{equation}
    where $e_p \sim \mathcal{N}(1, s_p)$. For neutron reflectometry, we apply constant instrumental resolution $\frac{\delta q}{q} = 5 \%$. 

    \paragraph{Training data} The models are trained on $300000$ mini-batches sampled according to the introduced training scheme. Each mini-batch contains $8192$ reflectometry curves, resulting in $N \approx 2.5 \cdot 10^{9}$ training samples. Data generation is performed during the training for every batch without repetition. In this way, the model cannot overfit on a fixed training dataset, further increasing the reliability of the solution. The training process takes approximately 30 hours using a single NVIDIA V100 GPU.

\subsection{Inference pipeline}\label{sec:methods:inference}

    During inference, the measured reflectometry data and the prior parameters are supplied to the PANPE model. Given the desired effective sample size ESS, the model provides the respective number of parameter samples $\{\bm\theta_i\}_{i=1}^N$, refined by either providing importance weights $r_i$ (PANPE-IS) or by running MCMC (PANPE-MCMC).

    In the following, we discuss the pre- and post- processing stages of the inference, as well as the model architecture. 

    \paragraph{Input pre-processing} The input to the network is a measured reflectivity data $\bm{R} = \{q_p, R(q_p), s_p\}_{p=1}^{n_q}$ and the set of prior parameters $\bm\phi = \{\theta^{\mathrm{min}}_j, \theta^{\mathrm{max}}_j\}_{j=1}^{n}$. The input data is therefore $3n_q + 2n$-dimensional, where $n_q$ is arbitrary. We first preprocess it as follows. 

    First, we apply the equivariant transformations discussed in Methods \ref{sec:methods:invariance} to standardize the data before inference. That includes calculating the scaling coefficient $u = q_{\mathrm{max}} / q_{\mathrm{exp}}$ to rescale the $q$ axis of the measured data to match the training $q$ range. The input prior parameters are transformed accordingly. After transforming the prior parameters and the $q$ axis, both the reflectometry curve $R(q)$ and the measurement uncertainties $s(q)$ are pre-processed using a logarithmic transformation $0.1 \cdot \log_{10}(R_q + 10^{-10}) + 0.5$. Finally, the prior parameters $\bm\phi_{\bm{scaled}}$ are normalized with respect to the absolute parameter ranges. 

    \paragraph{Embedding network}  We use an embedding network to convert the input data to a fixed-dimensional latent vector, which is then supplied to the normalizing flow model. We note that our embedding architecture should have an ability to handle input data of varying sizes. Our tests suggest that for a fixed discretization, convolutional neural networks (CNN) provide the best performance on reflectometry data among different architectures. To this end, for arbitrary discretization, we implement a trainable neural kernel, which acts as an intermediary step, adapting the data before it reaches the CNN. The kernel is a neural network $K(q_k, q, R_q, s_q)$ that ``interpolates'' reflectivity levels and the measurement uncertainties to a set of predefined points $\{ q_k \}_{k=1}^{n_k}$. We utilize three such kernels, featuring 16, 32, and 64 equidistant points, respectively. The spacing between these points defines the kernel's ``window''. The kernel outputs are averaged for $q$ points falling within this window. Each kernel is a multilayer perceptron with GELU activation functions and 4-channel input, a hidden layer with 32 channels, and a 2-channel output layer. Each of three kernels is coupled with a convolutional network discussed below. We note that this architecture is not supposed to be discretization-invariant, as the posterior can be highly sensitive to the choice of $q$ points in reflectometry.
    
    Each CNN is a sequence of 5 blocks, each block containing a 1D convolutional layer, followed by a batch-normalization layer and a GELU activation function. Convolutional layers features a kernel of size 3, stride$=2$, and padding$=1$. Consequently, the dimension of the processed data is (approximately) halved after each layer. The number of channels is doubled in each block, starting from 32 up to 512. 

    Outputs from three CNNs are concatenated with the preprocessed parameters $\bm\phi$ and provided to a multilayer perceptron. The final 256-dimensional latent representation of the input is supplied to the flow-based model.
    
    \paragraph{Flow-based model} A normalizing flow \cite{rezende2015variational} employs a series of reversible and differentiable transformations on a simple, \textit{base distribution} (in our case, the standard normal distribution). This approach generates a complex distribution from which samples can be efficiently drawn and evaluated. In this work, we employ a series of $40$ transformations, each transformation block being a composition of a coupling layer with monotonic rational-quadratic splines \cite{neuralsplineflows2019} and a batch normalization layer \cite{dinh2016density}. After each transformation block, the parameters are randomly permuted. 
    
    \paragraph{Refinement by likelihood-based methods} During the inference, we sample parameters $\bm\theta_i \sim p_{\mathrm{NN}}(\bm\theta | \bm{R})$ and generate parameters in batches with the corresponding log probabilities. The obtained curves are used for calculating importance sampling weights (PANPE-IS) and streaming estimation of sample efficiency $\eeff$. We continue this procedure until the effective sample size $\mathrm{ESS} = N \cdot \eeff$ reaches an adequate threshold which we set equal to 500. The same criterion is employed for the traditional importance sampling, where the prior distribution $p(\theta \newmid \bm\phi)$ is used as the proposal distribution.

    Alternatively, samples generated by PANPE are used as efficient initialization points for MCMC (PANPE-MCMC). In our work, we introduce GPU-accelerated PyTorch-based implementations of several Affine Invariant MCMC algorithms \cite{braak2006markov, Goodman2010, emcee2013Foreman} enabling near real-time MCMC-based refinement operation.

\subsection{Low sample efficiency estimation}\label{sec:methods:low-sample-efficiency}

    We consider two approaches for estimating sample efficiency for the conventional importance sampling method with prior acting as a proposal distribution. The first approach is the use of importance sampling weights via direct sampling from the prior distribution $p(\bm\theta)$:
    
    \begin{equation}\label{eq:eff-weights}
        \eeff = \frac{\langle w^{\mathrm{IS}}_i \rangle^2}{\langle\left(w^{\mathrm{IS}}_i\right)^2\rangle} \text{ ,}
    \end{equation}
    where $w^{\mathrm{IS}}_i = p(\bm{R} \newmid \bm\theta_i)$, $\bm\theta_i \sim p(\bm\theta)$, and $\langle \cdot \rangle $ is the average operation over all samples $i$.
    An accurate estimation requires $N = \mathrm{ESS} / \eeff$ samples, e.g. sample efficiency $\eeff = 10^{-12}$ requires more than $10^{12}$ samples, which is computationally unfeasible. An insufficient number of samples $N$ only provides an upper bound $\eeff < 1 / N$. Therefore, it can only be employed in practice for sufficiently high sample efficiencies.
    
    The alternative approach employs the analytical form \cite{MaiaPolo2022}:
    
    \begin{equation}\label{eq:eff-analytical}
         \eeff \xrightarrow{a.s.} \eeff^* =  \left(\mexpect_{\bm\theta \sim p(\bm\theta \newmid \bm{R})}\left[\frac{p(\bm\theta \newmid \bm{R})}{p(\bm\theta)}\right]\right)^{-1} = \frac{p(\bm{R})}{\mexpect_{\bm\theta \sim p(\bm\theta \newmid \bm{R})}\left[p(\bm{R} \newmid \bm\theta)\right]} \text{ .}
    \end{equation}
    
    In the case of the uniform prior distribution $p(\bm\theta)$, the equation simplifies to 
    
    \begin{equation}
          \left(\mexpect_{\bm\theta \sim p(\bm\theta \newmid \bm{R})}\left[\frac{p(\bm\theta \newmid \bm{R})}{p(\bm\theta)}\right]\right)^{-1} = \frac{v( p(\bm\theta \newmid \bm{R}))}{v( p(\bm\theta))} \text{ ,}
    \end{equation}
    where the quantity $v(p(\bm{x})) \equiv \left( \mexpect_{p(\bm{x})} \left[ p(\bm{x}) \right] \right)^{-1}$ can be interpreted as an ``efficient volume'' of the distribution $p(\bm{x})$. In this way, $v( p(\bm\theta)) = \prod_{j=1}^{n} (\theta_j^{max} - \theta_j^{min}) = \Theta$ is the volume of the prior distribution, and
    
    \begin{equation}\label{eq:posterior-volume}
        v(p(\bm\theta \newmid \bm{R})) = \left(\int_{\Theta} p(\bm\theta \newmid \bm{R})^2 d\theta\right)^{-1}
    \end{equation}
    characterizes the efficient volume of the posterior distribution. For instance, in the case of $d$-dimensional standard normal distribution $\mathcal{N}(0, \mathbbm{1} \cdot \sigma)$, $v(p) = (2\sqrt{\pi}\sigma)^d$. Naturally, the sample efficiency in our case is the ratio between the defined volume of the target distribution and the volume of the (uniform) proposal distribution.
    
    We estimate $\eeff^*$ using samples from our PANPE-IS model $\{\bm\theta_i\}_{i=1}^N$:
    \begin{equation}\label{eq:eff-estimation}
        \eeff^* = \frac{p(\bm{R})}{\mexpect_{\bm\theta \sim p(\bm\theta \newmid \bm{R})}\left[p(\bm{R} \newmid \bm\theta)\right]} \approx \frac{\left(\sum_{i=1}^N w_i\right)^2}{\sum_{i=1}^N w_i p(\bm{R} \newmid \bm\theta_i)} \text{ ,} 
    \end{equation}
    where the importance weights and samples provided by PANPE-IS should not be confused with the weights and samples from prior distribution in \autoref{eq:eff-weights}.
    
    In this way, we obtain $\epsilon_{\mathrm{IS}}$ estimations in the case of low sample efficiency of the IS method. However, when the prior distribution is narrow enough, we can estimate $\epsilon_{\mathrm{IS}}$ using both methods independently. SM Figure 7 illustrates the consistency between these two approaches.

\section{Acknowledgements}

    We acknowledge funding by the BMBF. This research was also supported in part by the Maxwell computational resources operated at DESY with the assistance of A. Rothkirch and F. Schlünzen. V. Starostin is grateful to S. Axen for insightful discussions. M. Dax wishes to thank J. H. Macke and B. Schölkopf for their support. F. Schreiber and Á. Tejero-Cantero are members of the Machine Learning Center of Excellence, EXC number 2064/1 – Project number 390727645.

\section{Data availability}

    All the experimental data used in the paper is publicly available.

\section{Code availability}

     The code implementing our method will be made publicly available at https://github.com/mlcolab/panpe (final link may change).

\section{Author contributions}

    V.S. conceived the concept for the work, implemented, trained, and tested neural networks, implemented and conducted the conventional Bayesian analysis. M.D. proposed the training strategy for prior amortization. M.D. and Á.T.-C. contributed expertise in machine learning. A.H., A.G., and F.S. contributed expertise in scattering. F.S. supervised the research process. All authors discussed the results and commented on the manuscript.

\section{Conflict of interests}

    The authors have no conflicts of interest to declare.

\bibliography{references}

\end{document}


\title{SUPPLEMENTARY INFORMATION \\
Fast and Reliable Probabilistic Reflectometry Inversion with Prior-Amortized Neural Posterior Estimation}

\author{Vladimir Starostin\textsuperscript{a}}
\author{Maximilian Dax\textsuperscript{b}}
\author{Alexander Gerlach\textsuperscript{a}}
\author{Alexander Hinderhofer\textsuperscript{a}}
\author{Álvaro Tejero-Cantero\textsuperscript{a}}
\author{Frank Schreiber\textsuperscript{a}}

\affiliation{
[a] University of Tübingen, 72076 Tübingen, Germany\\
[b] Max Planck Institute for Intelligent Systems, 72076 Tübingen, Germany
}

\maketitle
\newpage

\begin{figure}
    \centering
    \includegraphics[width=0.8\linewidth]{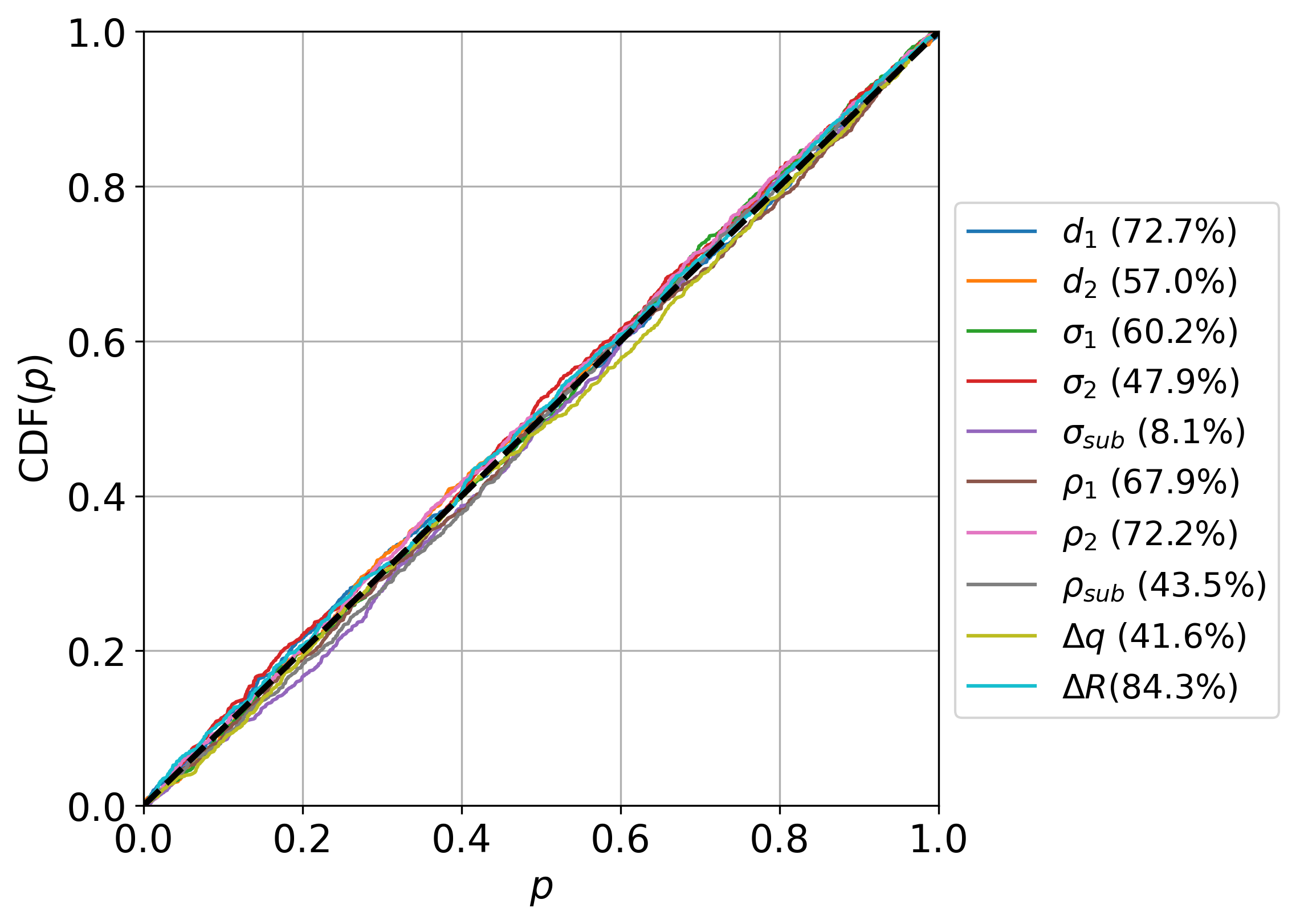}
    \caption{One-sample Kolmogorov–Smirnov tests of our PANPE model for 10 marginal distributions performed on 1000 simulated test curves, with $p$ values for individual parameters indicated in the legend. }
    \label{fig:ks-test}
\end{figure}

\begin{figure}
    \includegraphics[width=0.7\linewidth]{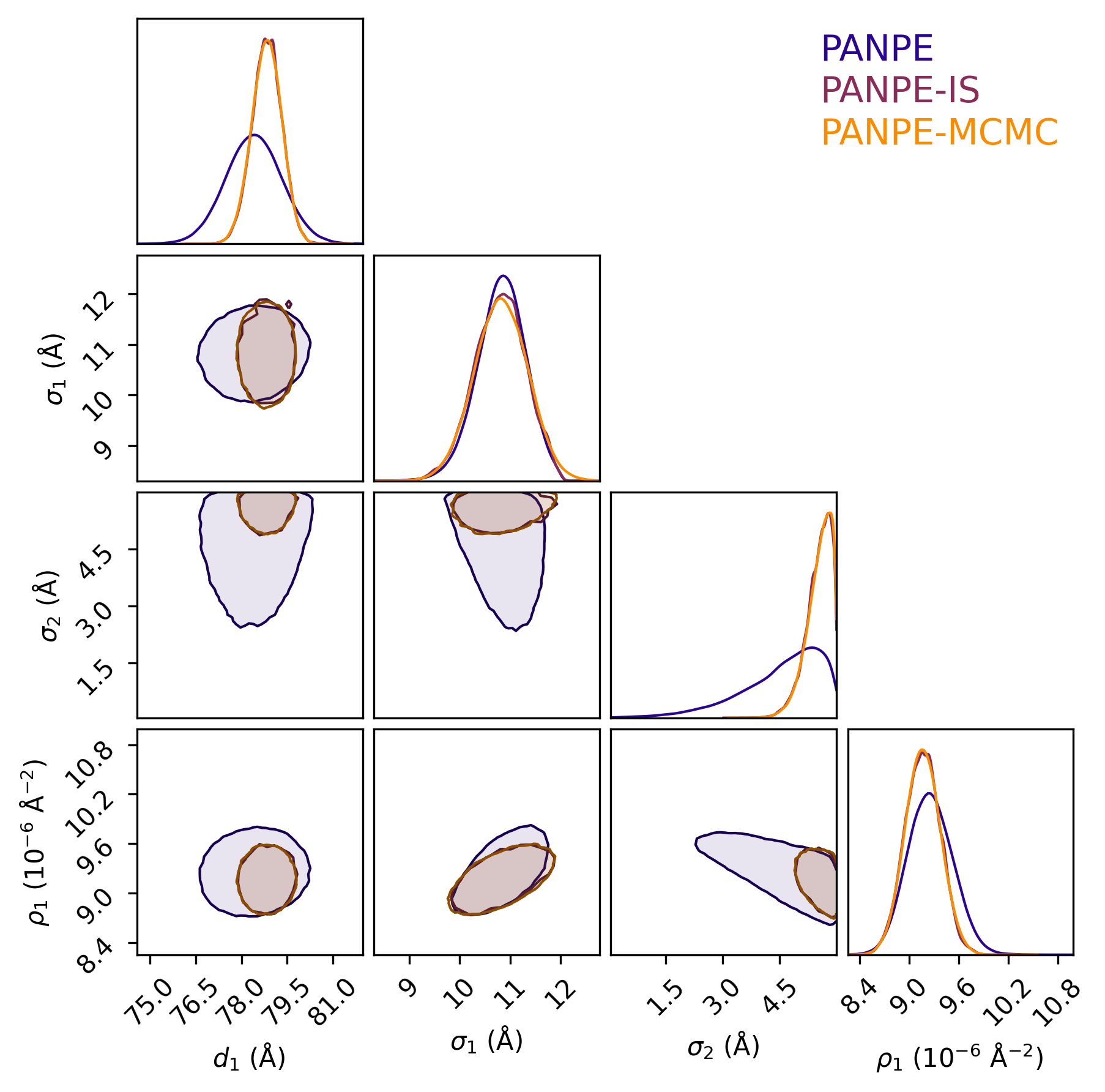}
    \caption{PANPE results on a measured XRR. The marginalized corner plot of 10-dimensional posterior distribution shows probability-mass covering proposal distribution provided by PANPE and two independently refined distributions via PANPE-IS and PANPE-MCMC, both of which yield equivalent solutions. }
    \label{fig:panpe-mcmc-snis-comparison}
\end{figure}

\begin{figure}
    \includegraphics[width=1\linewidth]{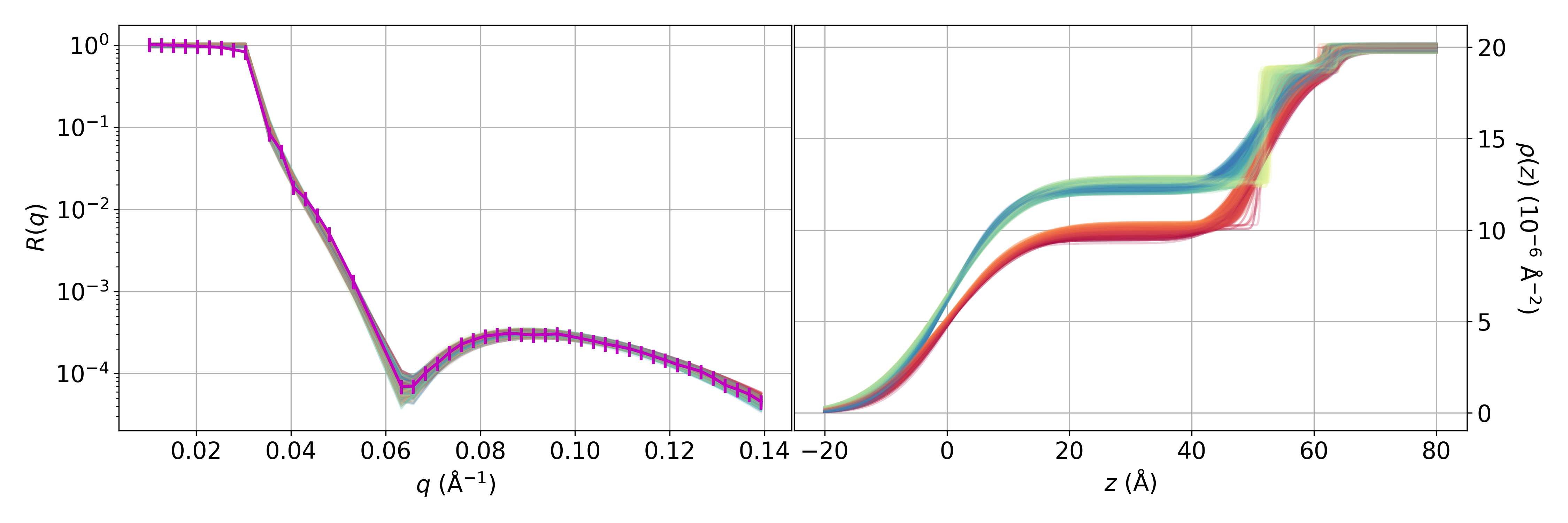}
    \caption{An example of a multimodal solution on a measured XRR curve. Reflectometry curves on the left include the measured curve, shown in magenta, alongside simulated curves corresponding to samples obtained through PANPE-IS, which colors match the colors of the respective SLD profiles on the right-hand side. The colors help recognise distinct SLD profiles. }
    \label{fig:exp-data-multimodal-example}
\end{figure}

\begin{figure}
    \centering
    \includegraphics[width=0.8\linewidth]{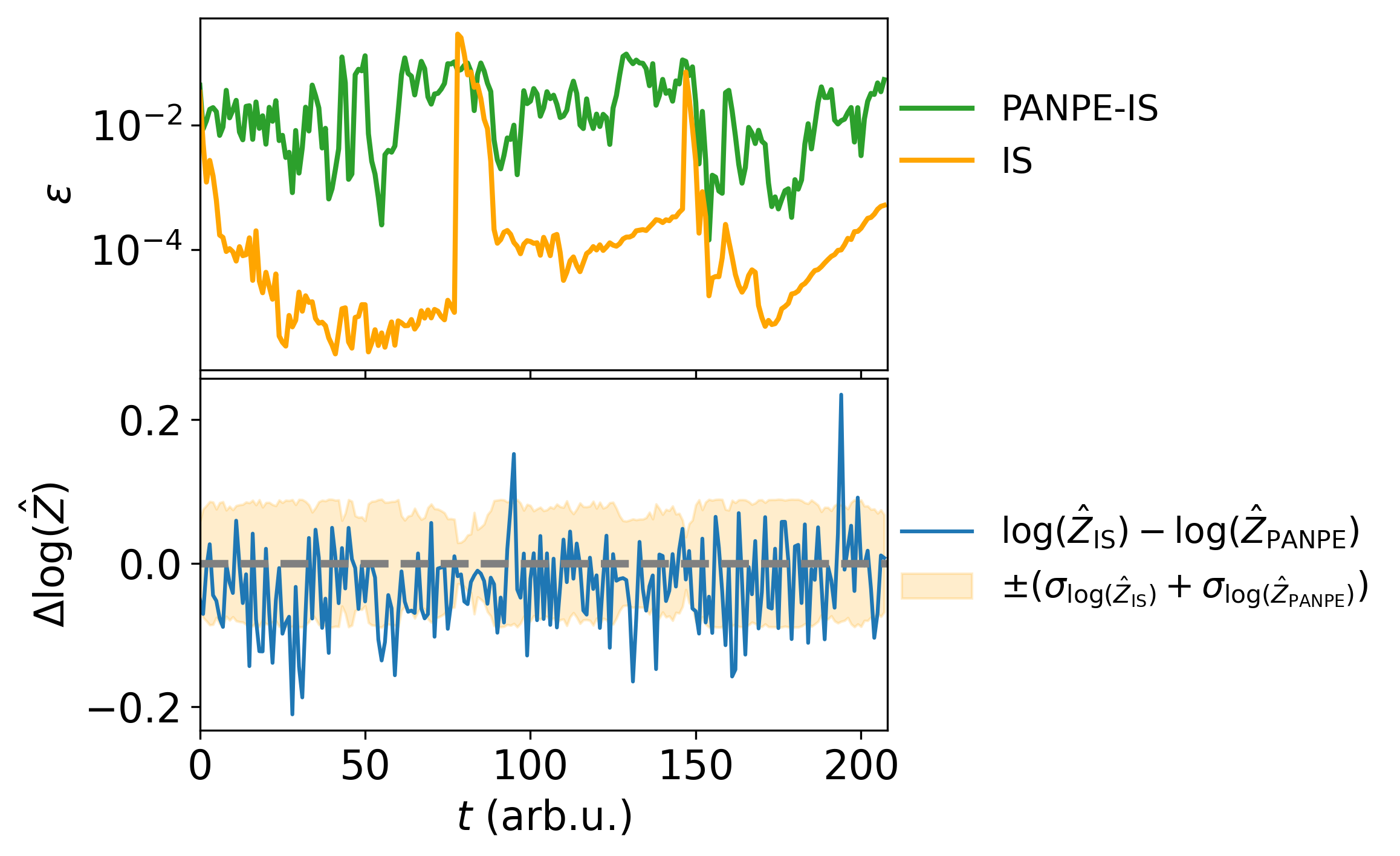}
    \caption{Top: sample efficiencies for conventional IS (orange) and PANPE-IS (green) on an experimental dataset of 208 X-Ray reflectometry curves. Here 3 datasets from real-time measurements are concatenated. Bottom: blue line is the difference between IS and PANPE-IS estimates of log evidence. The filled area denotes estimate errors. }
    \label{fig:performace-parallel-plot}
\end{figure}

\begin{figure}
    \centering
    \includegraphics[width=1\linewidth]{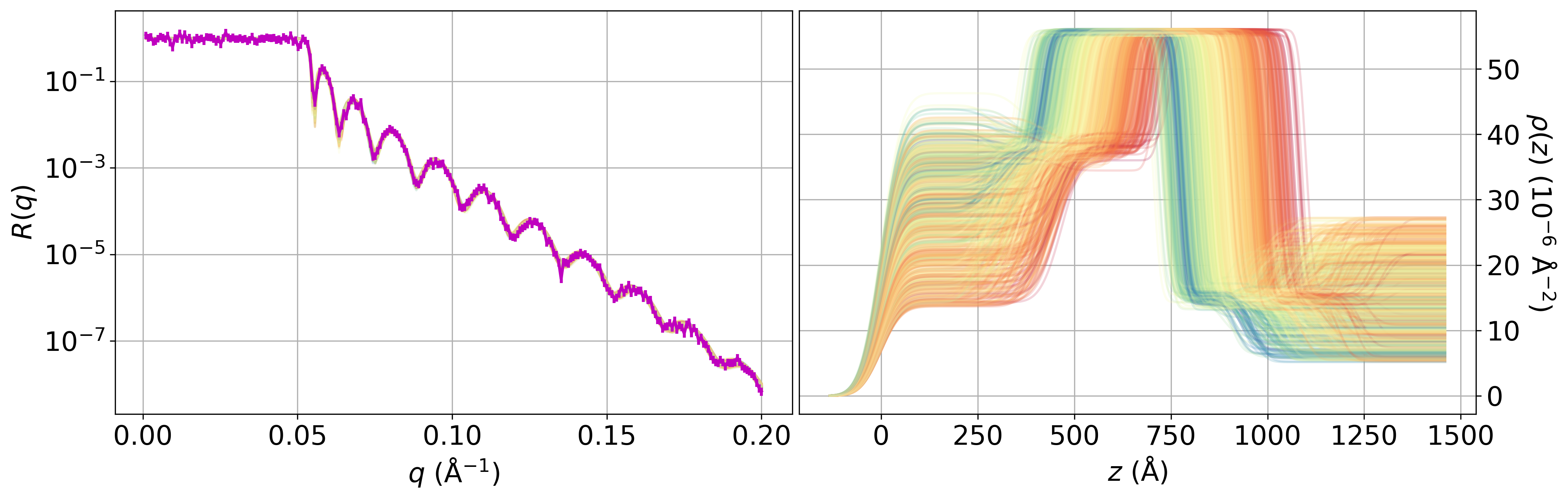}
    \caption{ The PANPE-IS inference result for a simulated four-layer structure, which includes 16 parameters, demonstrates that increasing the number of layers can lead to highly ambiguous results. This necessitates the use of more prior information or additional measurements.  }
    \label{fig:four-layers-example}
\end{figure}

\begin{figure}
    \centering
    \includegraphics[width=1\linewidth]{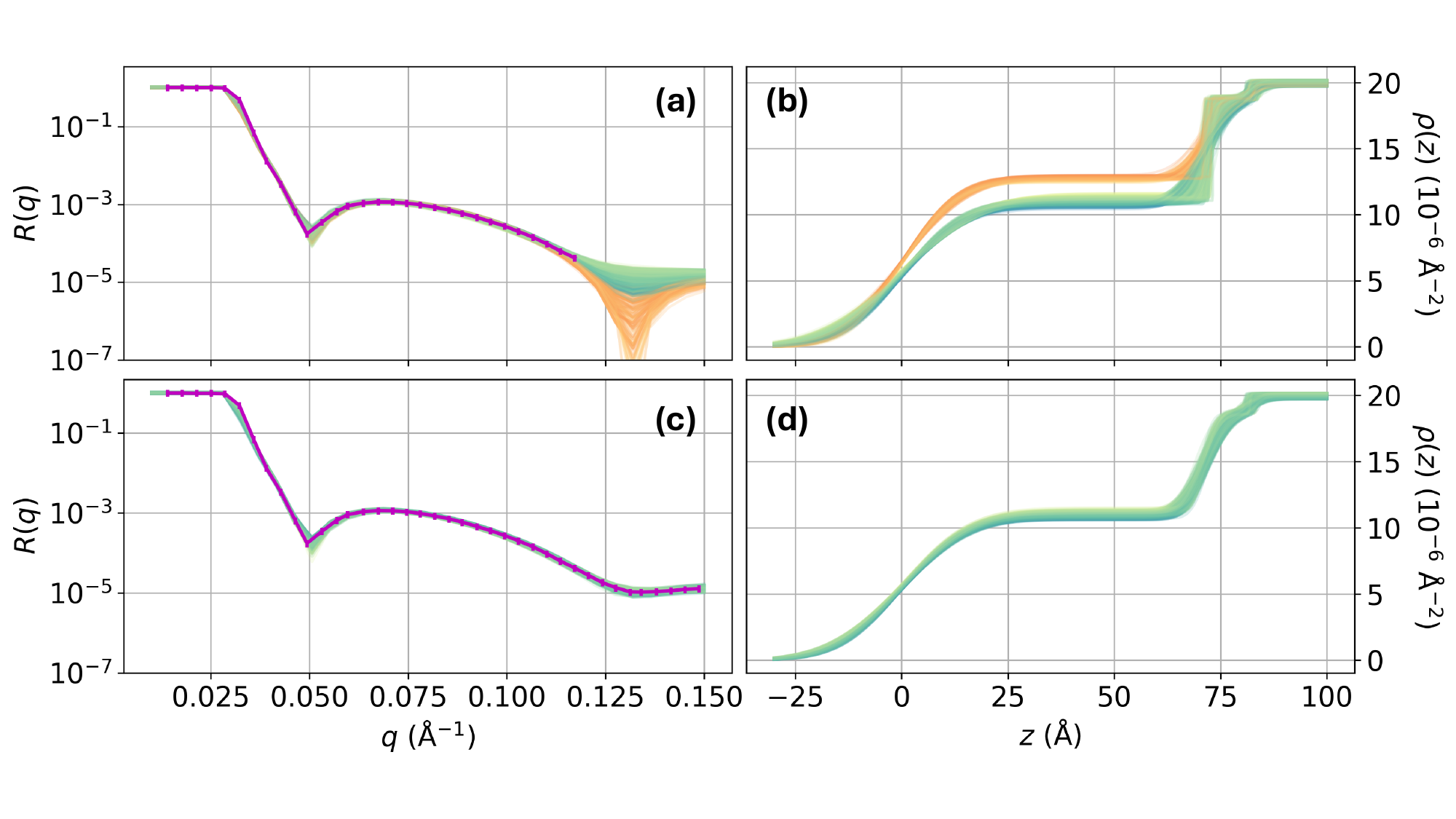}
    \caption{Analysis of an experimental XRR curve performed at two measurement times. (a) The experimental curve with 10 last $q$ points retrospectively removed to simulate the data status at an earlier time in the measurement. (c) The complete measured curve. (b) and (d) display the corresponding SLD profiles sampled using PANPE-IS. In cases of ambiguity, such as between (a) and (b), simulated curves can guide the measurement to focus on the most informative $q$ points (see the orange and green simulated curves in (a)). This approach can reduce the total number of required $q$ points and speed up the measurements. }
    \label{fig:q-plot}
\end{figure}

\begin{figure}
    \centering
    \includegraphics[width=0.6\linewidth]{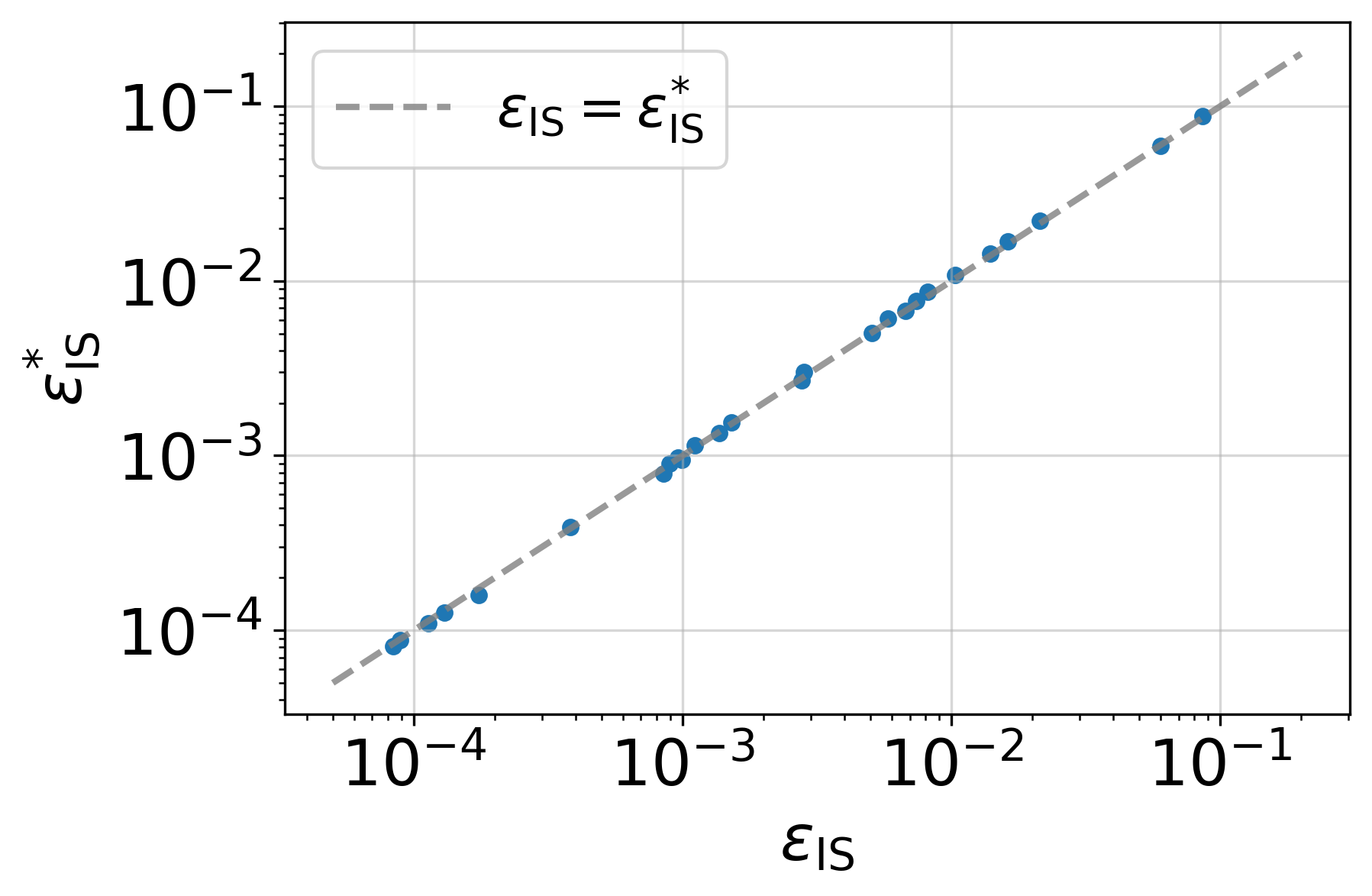}
    \caption{The consistency between the standard sample efficiency estimations of the conventional importance sampling ($\epsilon_{\mathrm{IS}}$ as per Equation 6) and via our model ($\epsilon^*_{\mathrm{IS}}$ as per Equation 10) calculated for simulated samples with sufficiently high efficiency $\epsilon_{\mathrm{IS}}$.}
    \label{fig:e_estar_consistency}
\end{figure}

\begin{figure}
    \centering
    \includegraphics[width=\linewidth]{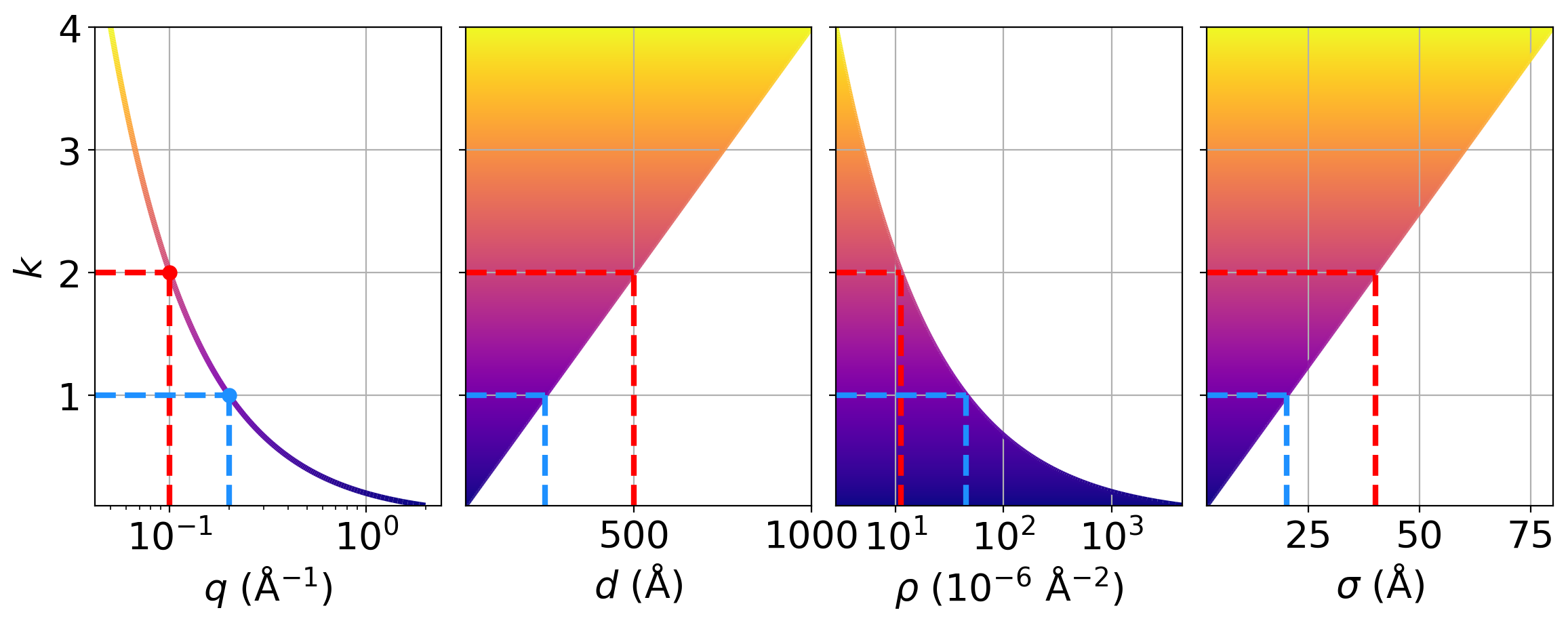}
    \caption{Scaling invariance of the SLD parameters and momentum transfer $q$. A model trained with a fixed $q$ range $q_{max} = 0.2$ \angstrom$^{-1}$ (blue dashed lines) can be then applied to the measured data with a different $q$ range, but the total parameter ranges are scaled accordingly. The red dashed lines demonstrate this scaling effect for $q_{max} = 0.1$ \angstrom$^{-1}$. The colored areas denote the ranges within which the trained model can be applied. }
    \label{fig:q-scaling}
\end{figure}